\documentclass[11pt,journal,onecolumn]{IEEEtran}

\pdfoutput=1

\usepackage{amsmath}
\usepackage{amsthm}
\usepackage{amsfonts}
\usepackage{upgreek}
\usepackage{amssymb}
\usepackage{setspace}
\usepackage{multicol}
\usepackage{multirow}
\usepackage[noadjust]{cite}
\usepackage{ctable}
\usepackage{array}
\usepackage{mdwmath}
\usepackage{mdwtab}
\usepackage{eqparbox}
\usepackage{url}
\usepackage{graphicx}
\usepackage{epsf}
\usepackage{mathrsfs}
\usepackage{cite}
\usepackage{dsfont}
\usepackage{epsfig}
\usepackage{enumerate}
\usepackage{hyperref}
\usepackage{cleveref}
\usepackage[caption=false]{subfig}

\newcolumntype{M}[1]{>{\centering\arraybackslash}m{#1}}

\newcommand{\subparagraph}{}
\usepackage{titlesec}
\titlespacing{\subsubsection}{\parindent}{2.25ex plus 1ex minus .2ex}{1.5ex plus .2ex}

\newtheorem{theorem}{Theorem}
\newtheorem{lemma}{Lemma}
\newtheorem{remark}{Remark}
\newtheorem{corollary}{Corollary}
\newtheorem{definition}{Definition}
\newtheorem{notation}{Notation}

\newtheorem{conjecture}{Conjecture}

\newcommand{\xc}{\mathbf{c}}
\newcommand{\xd}{\mathbf{d}}

\newcommand{\xh}{\mathbf{h}}

\newcommand{\xu}{\mathbf{u}}
\newcommand{\xv}{\mathbf{v}}

\newcommand{\xy}{\mathbf{y}}

\newcommand{\xA}{\mathbf{A}}
\newcommand{\xB}{\mathbf{B}}
\newcommand{\xC}{\mathbf{C}}

\newcommand{\xH}{\mathbf{H}}

\newcommand{\xR}{\mathbf{R}}

\newcommand{\xW}{\mathbf{W}}

\newcommand{\xomega}{\boldsymbol{\omega}}

\newcommand{\xxC}{\mathcal{C}}
\newcommand{\xxD}{\mathcal{D}}

\newcommand{\xxI}{\mathcal{I}}
\newcommand{\xxL}{\mathcal{L}}

\newcommand{\xxW}{\mathcal{W}}

\newcommand{\bbC}{\mathbb{C}}
\newcommand{\bbE}{\mathbb{E}}

\newcommand{\bbR}{\mathbb{R}}

\allowdisplaybreaks

\newcommand{\Def}{\triangleq}

\newcommand{\DoF}{\mathsf{DoF}}

\newcommand{\ie}{i.e., }

\usepackage{wasysym}
\newcommand{\phase}[1]{$\bullet$ \textbf{Phase #1}:}

\newcommand{\hop}[1]{$\triangleright$  \underline{\emph{Hop #1}}:}

\DeclareMathOperator*{\argmax}{arg\,max}
\DeclareMathOperator{\diag}{\textbf{diag}}

\crefname{lemma}{Lemma}{Lemmas}
\Crefname{lemma}{Lemma}{Lemmas}
\crefname{notation}{Notation}{Notations}
\Crefname{notation}{Notation}{Notations}
\crefname{remark}{Remark}{Remarks}
\Crefname{remark}{Remark}{Remarks}
\crefname{theorem}{Theorem}{Theorems}
\Crefname{theorem}{Theorem}{Theorems}
\crefname{definition}{Definition}{Definitions}
\Crefname{definition}{Definition}{Definitions}
\crefname{proposition}{Proposition}{Propositions}
\Crefname{proposition}{Proposition}{Propositions}
\crefname{conjecture}{Conjecture}{Conjectures}
\Crefname{conjecture}{Conjecture}{Conjectures}
\crefname{corollary}{Corollary}{Corollaries}
\Crefname{corollary}{Corollary}{Corollaries}
\crefname{section}{Section}{Sections}
\Crefname{section}{Section}{Sections}
\crefname{figure}{Fig.}{Figs.}
\Crefname{figure}{Figure}{Figures}
\crefname{table}{Table}{Tables}
\Crefname{table}{Table}{Tables}
\crefname{equation}{\hspace{-1mm}}{\hspace{-1mm}}
\Crefname{equation}{Equation}{Equations}

\long\def\symbolfootnote[#1]#2{\begingroup%
\def\thefootnote{\fnsymbol{footnote}}\footnote[#1]{#2}\endgroup}

\begin{document}
\title{Layered Interference Networks with Delayed CSI: DoF Scaling with Distributed Transmitters}
\author{\IEEEauthorblockN{Mohammad~Javad~Abdoli and~A.~Salman~Avestimehr}
\thanks{M. J. Abdoli is with Huawei Technologies Canada Co., Ltd., Ottawa, ON, Canada. Email: {\sffamily \scriptsize mjabdoli@gmail.com}. A. S. Avestimehr is with the School of Electrical and Computer Engineering, Cornell University, Ithaca, NY, USA. Email: {\sffamily \scriptsize avestimehr@ece.cornell.edu}.}
\thanks{This work was presented in part at the IEEE International Symposium on Information Theory (ISIT), 2013, Istanbul, Turkey \cite{abdoli2013multihopISIT}.}
\thanks{The work of A. S. Avestimehr is in part supported by NSF Grants CAREER-0953117, CCF-1161720, NETS-1161904, AFOSR Young Investigator Program Award, ONR award N000141310094, and a Qualcomm Gift.}
} 

\maketitle

\begin{abstract}
The layered interference network is investigated with delayed channel state information (CSI) at all nodes. It is demonstrated how multi-hopping can be utilized to increase the achievable degrees of freedom (DoF). In particular, a multi-phase transmission scheme is proposed for the $K$-user $2K$-hop interference network in order to systematically exploit the layered structure of the network and delayed CSI to achieve DoF values that scale with $K$. This result provides the first example of a network with distributed transmitters and delayed CSI whose DoF scales with the number of users.
\end{abstract}

\section{Introduction}

Interference management is a central challenge in the design and operation of wireless networks. To understand its fundamental limits, the $K$-user single-input single-output (SISO) interference channel\footnote{Throughout the paper, the terms ``network'' and ``channel'' refer to SISO configurations, unless explicitly stated otherwise.} 
has been a canonical example studied in multiuser information theory. For this network, the traditional and commonly-deployed approaches for interference management, such as interference orthogonalization, interference decoding, or treating interference as noise can achieve only one total degree of freedom (DoF). By using more elegant physical-layer interference management techniques, in particular interference alignment \cite{maddah2008communication, cadambe2008interference}, a total of $K/2$ DoF can be achieved in the $K$-user interference channel. This, at a coarse level, implies that, with appropriate design of physical-layer signaling, the total DoF of an interference network can \emph{scale} linearly with the number of users, despite the fact that all users communicate with each other over a shared wireless medium.

However, therein lies a critical problem. It is widely known that the perfect and instantaneous knowledge of channel state information (CSI) at transmitters (CSIT) plays a crucial role in achieving the full DoF promised by interference alignment. The perfect and instantaneous CSIT requires the capacity of the feedback links to scale with the network's size and also necessitates that the feedback delay be within the channel's coherence time. Therefore, in high-mobility environments, the instantaneous CSIT assumption is by no means realistic. As a result, recently, interest has been increasing in studying the impact of the lack of up-to-date CSIT in wireless networks.

In the context of broadcast channels, surprisingly, it was recently shown that even completely expired CSIT (a.k.a. delayed CSIT) yields DoF scaling \cite{maddah2012completely}. In particular, it was shown that the $K$-user multiple-input single-output (MISO) broadcast channel with at least $K$ antennas at transmitter has the following DoF under the delayed CSIT assumption.
\begin{align}
\frac{K}{1+\frac{1}{2}+\cdots+\frac{1}{K}}\underset{K\to \infty}{\approx} \frac{K}{\ln K} \label{Eq:BCDoF}
\end{align}
A key idea of \cite{maddah2012completely} in exploiting the delayed CSIT in broadcast channels was to align the past interference using a \emph{transmission-retransmission approach}. More specifically, a certain amount of information first is transmitted irrespective of the current CSI. Then, the \emph{entire} past interference at each unintended receiver is reconstructed by the transmitter using delayed CSI and the centralized knowledge of transmitted symbols, and this information is retransmitted to achieve the interference alignment. 

Several recent works have extended the aforementioned transmission-retransmission approach to interference channels. This includes $K$-user interference and X channels \cite{Maleki_Jafar_Retro,Abdoli2013IC-X,kang2013ergodic}, multi-antenna two-user interference and X channels \cite{Vaze2012IC_DCSIT,ghasemi2011interference,Ghasemi2011Xchannel,Ghasemi2012MIMOX}, and two-user binary fading interference channel \cite{vahid2011DNSI,vahid2013capacity}. Despite the remarkable gains offered by delayed CSI in broadcast channels, DoF benefits reported to date of delayed CSI over no CSI in the networks with distributed transmitters are quite marginal. In particular, although it has been shown that $K$-user interference and X channels can achieve more than one DoF with delayed CSIT \cite{Maleki_Jafar_Retro,Abdoli2013IC-X,kang2013ergodic}, the achieved DoF values are less than $2$ for any number of users in both channels. As further progress, in \cite{Abdoli2012FullDuplexFeedback} the transmitters were equipped with one-to-one output feedback together with delayed CSI, a.k.a. Shannon feedback. However, in spite of strict DoF improvements over the delayed CSIT case, their achieved DoF values for both $K$-user interference and X channels with Shannon feedback still do not scale with the number of users, and they are less than $2$ for any number of users.

A major challenge in attaining DoF improvements in interference channels with delayed CSIT is that, in such networks, the received interference is due to more than one transmitter, each of which has access \emph{only} to its own interference contribution. Therefore, per channel use of ``transmission phase,'' in order to align the received interference at a receiver, the number of interference contributions that must be retransmitted remains close to the number of active transmitters, which, in turn, is an upper bound on the number of independently-transmitted symbols. The latter is due to the fact that each transmitter has a single antenna. In other words, the number of to-be-retransmitted quantities per channel use follows the number of transmitted quantities closely. As a result, no scaling of DoF in interference channels with delayed CSIT has been achieved to date. This together with lack of non-trivial DoF upper bounds leaves the problem of DoF characterization of interference channels with delayed CSIT still open and challenging. It is even unknown whether or not the DoF of such networks \emph{scales} with the number of users.

In this paper, we explore the $K$-user layered interference networks with delayed knowledge of CSI at all nodes. This investigation was motivated by recent results that demonstrated that multi-hopping can significantly impact the DoF of interference networks with instantaneous and perfect CSI at all nodes by enabling new interference management strategies for two-unicast networks \cite{gou2011aligned,shomorony2011two-unicast} and for multi-unicast networks \cite{shomorony2012degrees}. It was previously shown in \cite{vaze2011degrees} that one layer of relays can increase the DoF to $4/3$ (as opposed to one) for $2$-user interference networks with delayed CSI at sources and no CSI at relays. In this paper, we particularly focus on $K\geq 3$ and seek to determine whether it is feasible to utilize multi-hopping in order to achieve DoF scaling with delayed CSI at all nodes. Although multi-hopping seems to be helpful in achieving a better communication performance, an inherent challenge here is that one must deal with a more intricate problem in the presence of several hops. 

We investigate the $K$-user $2K$-hop interference network with delayed CSI. By interpreting the original $2K$-hop interference network as a cascade of two $K$-hop X networks, we convert the original problem to the problem of communication over a $K$-user $K$-hop X network with delayed CSI. This provides the communication flexibility offered by the multi-broadcast nature of the X network while ensuring that the attained DoF is achievable in the original multi-unicast network. Then, for the $K$-user $K$-hop X network with delayed CSI, we propose a $K$-phase transmission scheme which, among its other ingredients, possesses two key ingredients, namely, \emph{symbol offloading} and \emph{hop-distributed partial scheduling and interference nulling (PSIN)}. Specifically, phase $m$, $1\leq m \leq K$, involves the transmission of order-$m$ symbols, which are of common interest of $m$ destination nodes and are available at layer-$(m-1)$ nodes. The role of the symbol offloading operation in phase $m$ is to transfer the order-$m$ symbols from layer-$(m-1)$ to layer-$m$ nodes. The key idea behind the proposed symbol offloading is that linear functions of the original order-$m$ symbols, rather than the original order-$m$ symbols themselves, are offloaded as \emph{new} order-$m$ symbols. While offloading the original order-$m$ symbols themselves is equivalent to transmission over a single-hop X network with delayed CSI, the proposed offloading is accomplished at the maximum DoF of $K$. Then, the symbol offloading is followed by the PSIN operation, which is performed by layer-$m$ nodes and aims at an ``interference-controlled'' transmission of the offloaded order-$m$ symbols. In particular, the role of PSIN in hop $m$ is twofold. First, it controls the number of interferers that contribute to each linear combination obtained by each layer-$(m+1)$ node. This is achieved jointly by an appropriate transmitter/destination scheduling as well as a redundancy transmission which enables each layer-$(m+1)$ node to partially null the received interference. Second, it enables the generation of order-$(m+1)$ symbols eventually at destination nodes by yielding linear combinations that contain an appropriate mixture of the order-$m$ symbols. Then, the linear combinations obtained by layer-$(m+1)$ nodes are forwarded to the destination nodes by amplify-and-forward operations of the subsequent layers.

As a surprising result, we show that the achievable DoF of the proposed transmission scheme scales with $K$, by showing that it grows asymptotically as fast as $\frac{1}{2}f^{-1}(K)$, where $f^{-1}$ is the inverse function of $f(x)\Def x^x$. While this achievable DoF scales very slowly with the number of users, the importance of this result is that it can be considered as the first example of a single-antenna network with distributed transmitters wherein the delayed CSI yields DoF scaling with the number of users. Since the gap between our achievable DoF and the best known upper bound, \ie the DoF of the $K$-user MISO broadcast channel given by \cref{Eq:BCDoF}, also scales with $K$, a new open problem arises concerning whether or not the achieved scaling rate is tight.

Further, we focus on the $3$-user case and show that our general $K$-user approach can be improved for $K=3$. In particular, we show that the $3$-user $2$-hop interference network can achieve $16/11\approx 1.454$ DoF with delayed CSI, as compared to the best known achievable DoF for the $3$-user (single-hop) interference channel, \ie $6/5=1.2$ \cite{kang2013ergodic}. Note that the DoF of the $3$-user multi-hop interference network is bounded above by $18/11\approx1.636$, which is the DoF of the $3$-user MISO broadcast channel with delayed CSIT \cite{maddah2012completely}.

In a practical wireless network, it is notable that the DoF notion is valid only over a specific SNR range. This is due to the multi-cluster nature of wireless networks and existence of inevitable out-of-cluster interference \cite{Lozano2013Fundamental}. However, in this paper, we consider a single-cluster network and investigate how multi-hopping can help to manage the interference in networks with distributed transmitters when CSI is delayed. In this regard, DoF is used as a metric to analyze the proposed interference management ideas and to differentiate them with state-of-the-art techniques. In other words, we use the DoF metric to indicate that our scheme involves new ingredients that are not present in the state-of-the-art interference management schemes.

The outline of the paper is as follows. The next section provides a setup for the investigated problem. \Cref{Sec:MainResults} presents our main results. \Cref{Sec:Overview} provides an overview of our transmission strategy for the $K$-user interference network with delayed CSI. The analytical details of the proposed scheme are provided in \cref{Sec:AnalyticalDetails}, and its achievable DoF is calculated in \cref{Sec:DoF_Analysis}. Our improved achievability result for the $3$-user multi-hop interference network is proven in \cref{Sec:3_user_multi_hop_interference}, and the paper is concluded in \cref{Sec:Conclusions}.

\section{System Setup}
\label{Sec:SystemSetup}
\begin{figure}
\centering
\includegraphics{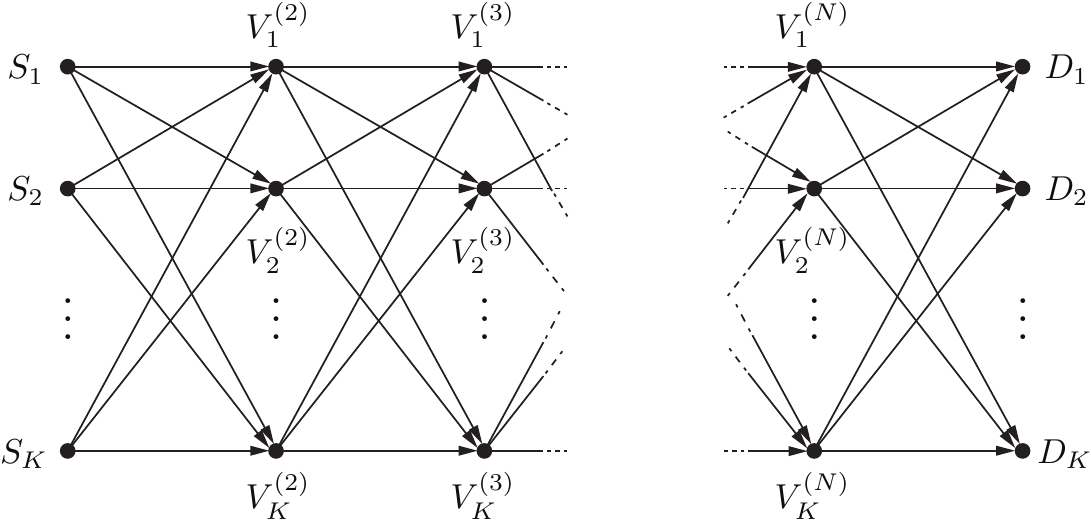}
\caption{$K$-user $N$-hop interference network.}
\label{Fig:NK_IC}
\end{figure}

A $K$-user $N$-hop interference network is defined as a set of $K$ source nodes denoted as $\{S_i\}_{i=1}^K$, a set of $K$ destination nodes denoted as $\{D_i\}_{i=1}^K$, and $N-1$ sets of intermediate nodes, called \emph{relays}, denoted as $\{V^{(n)}_i\}_{i=1}^K$, $2\leq n \leq N$. Also, $V^{(1)}_i$ and $S_i$ are used interchangeably throughout this paper, as are $V^{(N+1)}_i$ and $D_i$. There is a communication channel between each two consecutive layers of nodes in the network, called a hop, as depicted in \cref{Fig:NK_IC}. Each relay node operates in full-duplex mode, \ie it can transmit and receive simultaneously\footnote{Any achievable rate in the full-duplex multi-hop network is also achievable in the half-duplex network, at least with a factor of $1/2$.}. During time slot $t$, in hop $n$, node $V^{(n)}_j$ transmits $x^{(n)}_j(t)\in \bbC $ and node $V^{(n+1)}_i$ receives $y^{(n)}_i(t)\in \bbC$, where
\begin{align}
y^{(n)}_i(t) = \sum_{j=1}^K h^{(n)}_{ij}(t)x^{(n)}_j(t) + z^{(n)}_i(t), \quad 1\leq i \leq K,\quad 1\leq n \leq N,
\end{align}
and $h^{(n)}_{ij}(t)\in \bbC$ is the channel coefficient between $V^{(n)}_j$ and $V^{(n+1)}_i$, and $z^{(n)}_i(t)$ is the zero-mean unit-variance additive complex Gaussian noise at the input of $V^{(n+1)}_i$. The noise terms and channel coefficients are assumed to be independent and identically distributed (i.i.d.) over time and nodes. Moreover, the channel coefficients are assumed to be drawn according to a continuous distribution. Transmission in each hop is done over a block of $\tau$ time slots. The transmitted signal of each node is subject to the average power constraint $P$, \ie 
\begin{align}
\frac{1}{\tau}\sum_{t=1}^\tau \bbE |x^{(n)}_i(t)|^2 \leq P, \quad 1\leq i \leq K, \quad 1\leq n \leq N.
\end{align}

We assume that each source node $S_i$ has a message $W_i\in \xxW_i \Def \{1,2,\cdots, 2^{\tau R_i}\}$ of rate $R_i$ to communicate with its corresponding destination node $D_i$. We also consider a more general traffic setting, \ie the X network, in which each source node $S_i$ has a message $W_{ij}\in \xxW_{ij} \Def \{1,2,\cdots, 2^{\tau R_{ij}}\}$ of rate $R_{ij}$ to communicate with each destination node $D_j$. Denote the message set of source node $S_i$ and destination node $D_i$ by $\xxW^\textup{S}_i$ and $\xxW^\textup{D}_i$, respectively. Then, we have $\xxW^\textup{S}_i=\xxW^\textup{D}_i=\xxW_i$ in the interference network, and $\xxW^\textup{S}_i=\xxW_{i1}\times \cdots \times \xxW_{iK}$ and $\xxW^\textup{D}_i=\xxW_{1i}\times \cdots \times \xxW_{Ki}$ in the X network. Also, denoting the message vector of $S_i$ and $D_i$ by $\xW^\textup{S}_i$ and $\xW^\textup{D}_i$, respectively, we have $\xW^\textup{S}_i=\xW^\textup{D}_i=W_i$ in the interference network, and $\xW^\textup{S}_i=[W_{i1},W_{i2},\cdots,W_{iK}]^T$ and $\xW^\textup{D}_i=[W_{1i},W_{2i},\cdots,W_{Ki}]^T$ in the X network. Correspondingly, a rate tuple $\xR$ in the interference and X networks is defined as $\xR \Def [R_1,R_2,\cdots,R_K]^T$ and $\xR \Def [\xR^T_1, \xR^T_2,\cdots,\xR^T_K]^T$, respectively, where $\xR_i\Def [R_{i1},R_{i2},\cdots,R_{iK}]^T$ is the rate tuple of the source node $S_i$. 

Let us denote the CSI of the network in time slot $t$ by $\xH(t)$, which is a three-dimensional matrix of size $N\times K \times K$ that contains all channel coefficients of all hops in time slot $t$. In this paper, we consider a delayed CSI model in which $\xH(t)$ is assumed to be known at all nodes after a finite delay, which, for simplicity, is assumed to be one time slot\footnote{Assume that the delay in acquiring CSI is $D$ time slots. Then, by interleaving over $D$ blocks of communication, one can implement a transmission scheme that requires a delay of one time slot in CSI. Therefore, the only cost is a finite buffer of $D$ time slots at all nodes, which does not affect the capacity and, hence, the DoF.}. Then, we have the following definitions.

\begin{definition}[Block code with delayed CSI]
A block code of length $\tau$ and rate $\xR$ with delayed CSI is defined as a set of $K$ sequences of encoding functions
\begin{align}
\begin{split}
\varphi_{i,t,\tau}&: \xxW^\textup{S}_i \times \bbC^{NK^2(t-1)} \to \bbC \\
x_i^{(1)}(t)&=\varphi_{i,t,\tau}\left(\xW^\textup{S}_i,\left\{\xH(t')\right \}_{t'=1}^{t-1}\right)
\end{split},\quad\quad 1\leq t \leq \tau,\quad 1\leq i \leq K,
\end{align}
$(N-1)K$ sequences of relaying functions
\begin{align}
\begin{split}
\rho^{(n)}_{i,t,\tau}&: \bbC^{t-1} \times \bbC^{NK^2(t-1)}\to \bbC \\
x_i^{(n)}(t)&=\rho^{(n)}_{i,t,\tau}\left(\left\{y^{(n-1)}_i(t'),\xH(t')\right \}_{t'=1}^{t-1}\right)
\end{split},\quad\quad 1\leq t \leq \tau,\quad 1\leq i \leq K,\quad 2\leq n\leq N,
\end{align}
and $K$ decoding functions
\begin{align}
\begin{split}
\psi_{i,\tau}& :\mathbb{C}^{\tau}\times \bbC^{NK^2\tau} \to \xxW^\textup{D}_i \\
\hat{\xW}^\textup{D}_i&=\psi_{i,\tau}\left(\left\{y^{(N)}_i(t),\xH(t)\right\}_{t=1}^\tau\right)
\end{split},\quad\quad 1\leq i \leq K.
\end{align}
\end{definition}

\begin{definition}[Probability of error] The probability of error of a block code $C_\tau$ of length $\tau$ with delayed CSI is defined as
\begin{align}
P_e(C_\tau)\Def \textup{Pr} \left \{\bigcup_{i=1}^K \left\{\xW^\textup{D}_i\neq\psi_{i,\tau}\left(\left\{y^{(N)}_i(t),\xH(t)\right\}_{t=1}^\tau\right)\right\} \right \},
\end{align}
where $\xW^\textup{D}_i$ is the message vector transmitted for $D_i$.
\end{definition}

\begin{definition}[Achievable rate and capacity region] For a given power constraint $P$, a rate tuple $\xR(P)$ is said to be achievable in the $K$-user $N$-hop network with delayed CSI if there exists a sequence $\{C_\tau\}_{\tau=1}^\infty$ of block codes with delayed CSI, each with rate $\xR(P)$, such that $\lim_{\tau \to \infty}P_e(C_\tau)=0$. The closure of the set of all achievable rate tuples is called the capacity region with delayed CSI, and it is denoted by $\xxC(P)$.
\end{definition}

\begin{definition}[Degrees of freedom] The degrees of freedom (DoF) region of the $K$-user $N$-hop network with delayed CSI is defined as $\xxD\Def \lim_{P\to \infty} \frac{\xxC(P)}{\log_2 P}$. Any tuple $\xd\in \xxD$ is called an achievable DoF tuple and the summation of its elements is called an achievable sum-DoF, or simply achievable DoF. The supremum of all achievable sum-DoFs in the $K$-user $N$-hop interference (resp.\ X) network is called the network sum-DoF, or simply DoF, with delayed CSI, and it is denoted by $\DoF^\textup{IC}(K,N)$ (resp.\ $\DoF^\textup{X}(K,N)$).
\end{definition}

\section{Main Results}
\label{Sec:MainResults}

Our main result is the provision of an achievable DoF for the general multi-hop interference network with delayed CSI. More formally, we have the following theorem, which will be proven in \cref{Sec:Overview,Sec:AnalyticalDetails,Sec:DoF_Analysis}.

\begin{theorem}
\label{Th:2K_hop_IC_DoF}
The DoF of the $K$-user $2K$-hop interference network with $K\geq 3$ and delayed CSI satisfies
\begin{align}
\DoF^\textup{IC}(K,2K) \geq \frac{1}{t_1(q,K)+t_2(q,K)}, \label{Eq:DoF_lowerbound}
\end{align}
where
\begin{align}
t_1(q,K)&\Def \frac{1}{q-1}\left(\frac{\Gamma(q^{-1})(K-1)!}{\Gamma(K+q^{-1})}-\frac{1}{K}\right), \label{Eq:t_1_Def}\\
t_2(q,K)&\Def \frac{Kq+1}{q(q+1)K}+\frac{(2q-1)(K-1)}{2K[(K-1)q+1]}, \label{Eq:t_2_Def}
\end{align}
and $2\leq q \leq K-1$ is an arbitrary integer, and $\Gamma(\cdot)$ is the gamma function.
\end{theorem}

\begin{remark} It was conjectured in \cite{Abdoli2013IC-X} that the DoF of the $K$-user single-hop interference network with delayed CSI does not scale with the number of users. An important consequence of \cref{Th:2K_hop_IC_DoF} is that multi-hopping can provide DoF scaling in the $K$-user interference network with delayed CSI. Indeed, this can be considered as the first example of a network with distributed transmitters and delayed CSI whose DoF scales with the number of users. 
\end{remark}
In particular, we have the following corollary that is proven in Appendix \ref{App:ScalingProof}.
\begin{corollary}
\label{Cor:DoFScaling}
The DoF of the $K$-user $2K$-hop interference network with delayed CSI scales with $K$. Specifically, the following inequality provides an asymptotic lower bound to $\DoF^\textup{IC}(K,2K)$.
\begin{align}
\DoF^\textup{IC}(K,2K) \geq \frac{1}{2}f^{-1}(K)(1-\delta_K),
\end{align}
where $f^{-1}$ is the inverse function of $f(x)\Def x^x$, and $\delta_K>0$ goes to zero as $K\to \infty$.
\end{corollary}

\begin{remark}
Although \cref{Cor:DoFScaling} demonstrates DoF scaling of the multi-hop interference network, the achieved scaling rate is quite slow. Without a matching upper bound, the problem of characterizing the DoF scaling of this network with delayed CSI remains open.
\end{remark}

\begin{remark}
\label{Remark:3user6hop}
The lower bound of \cref{Th:2K_hop_IC_DoF} is not tight. For instance, while \cref{Th:2K_hop_IC_DoF} achieves $15/11$ DoF\footnote{Inequality \eqref{Eq:DoF_lowerbound} provides DoF lower bound of $1/(\frac{53}{90}+\frac{11}{15})\approx 0.756$ for the $3$-user $6$-hop interference network. However, as the proof of \cref{Th:2K_hop_IC_DoF} in \cref{Sec:3user3hop} shows, the proposed scheme achieves $1/\max\{\frac{53}{90},\frac{23}{45},\frac{11}{15}\}=\frac{15}{11}$ DoF for this network.} for the $3$-user $6$-hop interference network, the following theorem shows that this achievable DoF can be improved.
\end{remark}
\begin{theorem}
\label{Th:3_user_2_hop_IC_DoF}
The DoF of the $3$-user $2$-hop interference network with delayed CSI satisfies the following inequality.
\begin{align}
\frac{16}{11}\approx 1.454 \leq \DoF^\textup{IC}(3,2) \leq \frac{18}{11}\approx1.636. \label{Eq:3_user_2_hop_lower}
\end{align}
\end{theorem}
The upper bound of $18/11$ is indeed the DoF of the $3$-user MISO broadcast channel with delayed CSI \cite{maddah2012completely}. The proof of the lower bound is provided in \cref{Sec:3_user_multi_hop_interference}.

\begin{remark}
It is known that the $3$-user single-hop interference network in the i.i.d. fading environment has $1.5$ DoF with instantaneous CSI \cite{cadambe2008interference} and no more than $1$ DoF without CSI at the transmitters \cite{vaze2012NoCSIT}. Also, the best known achievable DoF for this single-hop network with delayed CSIT is $6/5 = 1.2$ \cite{kang2013ergodic}
\end{remark}

\section{Overview of the Transmission Strategy for the $K$-user Network}
\label{Sec:Overview}

In order to prove \cref{Th:2K_hop_IC_DoF}, we propose a multi-phase transmission scheme that achieves the lower bound of \cref{Eq:DoF_lowerbound}. In this section, we present an overview of our transmission strategy and highlight the key ideas on which the main building blocks of our transmission scheme are based. The analytical details of the scheme are provided in \cref{Sec:AnalyticalDetails}.
\subsection{Cascaded X Network Approach}
\label{Sec:X_Approach}

We consider the $K$-user $2K$-hop interference network as a cascade of two $K$-user $N$-hop X networks. Then, we have the following lemma.
\begin{lemma}
\label{Lem:X_Approach}
Any symmetric achievable DoF in the $K$-user $N$-hop X network is also achievable in the $K$-user $2N$-hop interference network that is formed by cascading two copies of the X network.
\end{lemma}
\begin{IEEEproof}
Assume that $D$ DoF is achievable in the $K$-user $N$-hop X network with $\{V^{(N)}_i\}_{i=1}^K$ as its destination nodes. Using its corresponding achievability scheme, one can transmit $K^2$ information symbols, \ie one symbol per source-destination pair, over the X network in $K^2/D$ time slots. Let us denote the information symbols of $S_j$ in the $2N$-hop interference network (which are all desired by $D_j$) by $u^{[j]}_1,\cdots, u^{[j]}_K$. Using the X network transmission scheme, it takes $K^2/D$ time slots to deliver $\{u^{[1]}_k,\cdots,u^{[K]}_k\}_{k=1}^K$ to $\{V^{(N)}_k\}_{k=1}^K$. Then, using the same scheme in the cascaded $N$-hop X network, it takes $K^2/D$ time slots to deliver $\{u^{[j]}_1,\cdots,u^{[j]}_K\}_{j=1}^K$ to $\{D_j\}_{j=1}^K$. Therefore, the whole $2N$-hop interference network spends $K^2/D$ time slots to deliver the $K^2$ symbols $\{u^{[j]}_1,\cdots,u^{[j]}_K\}_{j=1}^K$ to $\left\{D_j\right\}_{j=1}^K$. This implies achievability of $D$ DoF in the $2N$-hop interference network.
\end{IEEEproof}

According to the above lemma, it is sufficient to propose a transmission scheme that achieves the lower bound of \cref{Eq:DoF_lowerbound} in the $K$-user $K$-hop X network. The rest of this section provides an overview of our scheme for this network.

\subsection{Overview of the $K$-phase Transmission Scheme for the $K$-user $K$-hop X Network}
\label{Sec:Kphase_Overview}
The transmission scheme has $K$ distinct phases that operate sequentially, starting with phase $1$. As summarized in \cref{Tbl:Operations}, each phase involves a subset of hops, from a specific hop on to hop $K$. In phase $1$, the information symbols (order-$1$ symbols) are fed to the network. During phase $m$, $1\leq m \leq K-1$, so-called order-$m$ symbols are transmitted over the network and order-$(m+1)$ symbols are generated such that, if all of the generated order-$(m+1)$ symbols are delivered to their respective $(m+1)$-tuples of destination nodes, then all of the transmitted order-$m$ symbols will become resolvable by their respective $m$-tuples of destination nodes. Phase $K$ is responsible for delivering order-$K$ symbols to all the destination nodes. 

\begin{table*}
\caption{Operations of different hops in the $K$-phase transmission scheme for the $K$-user $K$-hop X network}
{\small
\begin{center}
\begin{tabular}{M{0.9cm}M{1.8cm}M{1.6cm}M{1.4cm}M{.9cm}M{1.8cm}M{1.8cm}M{3cm}}
\toprule
Phase & Hop $1$ & Hop $2$ & Hop $3$ & $\cdots$ & Hop $K-2$ & Hop $K-1$ & Hop $K$\\
\midrule
\midrule
$1$ & PSIN & AF & AF & $\cdots$ & AF & AF & generation of order-$2$ symbols\\
\midrule
$2$ & symbol offloading & PSIN & AF  & $\cdots$ & AF & AF & generation of order-$3$ symbols\\
\midrule
$3$ & silent & symbol offloading & PSIN  & $\cdots$ & AF & AF & generation of order-$4$ symbols\\
\midrule
\\
$\vdots$ & $\vdots$ & &  & $\ddots$ &  & & $\vdots$ \\
\\
\midrule
$K-1$ &  silent &  silent &  silent &   silent & symbol offloading & PSIN & generation of order-$K$ symbols\\
\midrule
$K$ &   silent &  silent &  silent &   silent & silent & symbol offloading & final delivery\\
\bottomrule
\end{tabular}
\end{center}}
\label{Tbl:Operations}
\end{table*}

Let us present the formal definition of an order-$m$ symbol.

\begin{definition}[Order-$m$ Symbol]
\label{Def:Order_m}
Order-$1$ symbols are defined as original information symbols. For any $2\leq m \leq K$, an order-$m$ symbol is defined as a piece of information that
\begin{itemize}
\item is intended to be delivered to a subset of $m$ destination nodes
\item is available at a source or relay node in the network
\end{itemize}
\end{definition}
As indicated in \cref{Tbl:Operations}, the proposed scheme is built upon the following key ideas. We postpone the analytical details of the scheme to \cref{Sec:AnalyticalDetails}.

\subsubsection{Hop-distributed partial scheduling and interference nulling (PSIN)}
\label{Sec:OverviewPSIN}
Except for phase $K$, which is responsible for delivering order-$K$ symbols to all destination nodes, each phase of the scheme involves the transmission of pieces of information that are desired by a subset of destination nodes. This implies that a potential interference for the unintended destination nodes is contained in the information flow passing through the hops during each phase $m$, $1\leq m \leq K-1$. In order to control the multi-interferer nature of the interference, we propose a technique called hop-distributed PSIN. In particular, the PSIN task, which is entrusted to one of the hops (PSIN hop) during each phase, has the following main ingredients.
\begin{itemize}
\item \emph{Partial transmitter scheduling}: By scheduling a subset of cardinality $L$ out of the $K$ transmitting source/relay nodes of the PSIN hop to transmit per time slot, the number of potential interferers is partially controlled, where $L<K$ is a design factor. One should note that, by such a scheduling mechanism, no more than $L$ DoF can be achieved by the transmission scheme. Therefore, in order to achieve DoF scaling, $L$ must scale with $K$. 
\item \emph{Partial destination scheduling}: By scheduling a subset of cardinality $m$ out of the $K$ destination nodes in the PSIN hop (\ie hop $m$) of phase $m$, the $m$ scheduled destination nodes will eventually receive interference-free linear combinations by the end of this phase.
\item \emph{Partial interference nulling}: The $L$ scheduled transmitting nodes of the PSIN hop transmit some redundancy together with the order-$m$ symbols so that the receiving relay nodes are enabled to null out the effect of one of the $L$ interferers from the signals they receive.
\end{itemize}

An important observation here is that due to the transmitter scheduling of the PSIN hop, it is one of the main resource-consuming hops in each phase. By distributing the PSIN task over the network hops, we ensure that each hop is responsible for the PSIN task in no more than one phase (\cref{Tbl:Operations}).
\subsubsection{Symbol offloading}
\label{Sec:OverviewOffloading}
The hop-distributed PSIN requires that the layer-$m$ nodes $\{V^{(m)}_i\}_{i=1}^K$ have access to the order-$m$ symbols during phase $m$, \ie each order-$m$ symbol is available at one of these nodes. Assuming $\{V^{(m-1)}_i\}_{i=1}^K$ have had access to the order-$(m-1)$ symbols in phase $m-1$, they would also have access to the order-$m$ symbols by the end of phase $m-1$. This is true since each order-$m$ symbol is, by construction, a function of order-$(m-1)$ symbols and past CSI. By symbol offloading, $\{V^{(m-1)}_i\}_{i=1}^K$ offload the order-$m$ symbols to $\{V^{(m)}_i\}_{i=1}^K$ during phase $m$. The key idea here is that linearly transformed versions of the original order-$m$ symbols, rather than the original order-$m$ symbols themselves, are offloaded as \emph{new} order-$m$ symbols to the next layer relays. While offloading the original order-$m$ symbols themselves is equivalent to transmission over a single-hop X network with delayed CSI for which the best known achievable DoF is less than $2$ \cite{Abdoli2013IC-X}, the proposed symbol offloading is accomplished at $K$ symbols per time slot.
\subsubsection{Hop silencing}
\label{Sec:OverviewSilencing}
As another consequence of the proposed symbol offloading, the transmission in phase $m\geq 2$ involves only hops $m-1$ to $K$. In other words, hops $1$ to $m-2$ are silent during phase $m$. This, in conjunction with the hop-distributed PSIN, helps in distributing the transmission load over the network hops.
\subsubsection{Amplify-and-forward (AF)}
\label{Sec:OverviewAF}
During phase $m\leq K-1$, the nodes $\{V^{(m+1)}_i\}_{i=1}^K$ amplify-and-forward the information obtained via the PSIN hop. Subsequently, $\{V^{(\ell)}_i\}_{i=1}^K$, $m+2\leq \ell \leq K$, perform AF operations to pass the information to the destination nodes. 

\subsubsection{Generation of order-$(m+1)$ symbols}
\label{Sec:OverviewOm+1}
By the end of phase $m$, the $m$ scheduled destinations receive some information about their desired order-$m$ symbols. However, the received information is not enough to decode the desired symbols. Retransmission of appropriate functions of the side information received by a non-scheduled destination node as order-$(m+1)$ symbols provides the $m$ scheduled destinations with the desired extra information and also aligns the past interference at the non-scheduled destination.

\section{Analytical Details of the Proposed Strategy for the $K$-user Network}
\label{Sec:AnalyticalDetails}

In this section, we provide analytical details of our transmission strategy for the $K$-user $K$-hop X network. To this end, we first elaborate on $K=3$ as an illustrative example. Then, we present the transmission strategy for the general $K$-user setting.

\subsection{$3$-user $3$-hop X Network}
\label{Sec:3user3hop}

For the $3$-user $3$-hop X network, the transmission scheme operates in three phases as outlined in \cref{Tbl:3User3hopX_Operations} and detailed in the following, and it achieves $15/11$ DoF.
\begin{notation}
We introduce the following notations.
\begin{itemize}
\item[] $u^{[i|j]}$: An information symbol of source $S_i$ for destination $D_j$
\item[] $T^{(k)}_m$: Time duration of hop $k$ in phase $m$ of the transmission scheme
\item[] $N_m$: Number of order-$m$ symbols that are transmitted during phase $m$ over the network
\end{itemize}
\end{notation}

\begin{table*}
\caption{Operations of different hops in the $3$-phase transmission scheme for the $3$-user $3$-hop X network}
{\small
\begin{center}
\begin{tabular}{M{0.9cm}M{3cm}M{3cm}M{5cm}}
\toprule
Phase & Hop $1$ & Hop $2$ & Hop $3$\\
\midrule
\midrule
$1$ & PSIN & AF & generation of order-$2$ symbols\\
\midrule
$2$ & symbol offloading & PSIN & generation of order-$3$ symbols\\
\midrule
$3$ &   silent & symbol offloading & final delivery\\
\bottomrule
\end{tabular}
\end{center}}
\label{Tbl:3User3hopX_Operations}
\end{table*}

\phase{$1$}

\hop{$1$ (PSIN)}

Let $L=3$ be fixed. In hop $1$, $18$ information symbols, all desired by a specific destination, say $D_1$, are transmitted in seven time slots as follows. During the first six time slots, each source node transmits a fresh information symbol in each time slot, \ie
\begin{align}
x^{(1)}_j(t) = u^{[j|1]}_t,\qquad 1\leq j \leq 3,\quad 1\leq t \leq 6. \label{Eq:3user3hop-PSIN-phase1_1}
\end{align}
In time slot $t=7$, each source node transmits the summation of its six previously-transmitted symbols.
\begin{align}
x^{(1)}_j(7) = u^{[j|1]}_1+u^{[j|1]}_2+\cdots+u^{[j|1]}_6, \quad 1\leq j \leq 3. \label{Eq:3user3hop-PSIN-phase1_2}
\end{align}
Hence, ignoring the noise, for any $1\leq i\leq 3$ we have
\begin{align}
y^{(1)}_i(t) &= \sum_{j=1}^{3}h^{(1)}_{ij}(t)u^{[j|1]}_t,\qquad \quad 1\leq t \leq 6, \label{Eq:3user3hop-yV-phase1_1}\\
y^{(1)}_i(7) &= \sum_{j=1}^{3}h^{(1)}_{ij}(7)\sum_{t=1}^6 u^{[j|1]}_t. \label{Eq:3user3hop-yV-phase1_2}
\end{align}

We note that each source node contributes exactly six information symbols to the seven received signals of each relay. Therefore, each relay can apply three different linear transformations on its seven received signals to obtain three different linear combinations, in each of which the contribution of one source node is nulled. In particular, relay $V^{(2)}_i$, $1\leq i \leq 3$, obtains the following linear combinations from \cref{Eq:3user3hop-yV-phase1_1,Eq:3user3hop-yV-phase1_2}.
\begin{align}
&\frac{y^{(1)}_i(7)}{h^{(1)}_{i1}(7)}-\sum_{t=1}^6 \frac{y^{(1)}_i(t)}{h^{(1)}_{i1}(t)}=L_{i\backslash1}(\xu^{[2|1]})+L_{i\backslash1}(\xu^{[3|1]}), \label{Eq:PSIN_Output1} \\
&\frac{y^{(1)}_i(7)}{h^{(1)}_{i2}(7)}-\sum_{t=1}^6 \frac{y^{(1)}_i(t)}{h^{(1)}_{i2}(t)}=L_{i\backslash2}(\xu^{[3|1]})+L_{i\backslash2}(\xu^{[1|1]}),\label{Eq:PSIN_Output2}  \\
&\frac{y^{(1)}_i(7)}{h^{(1)}_{i3}(7)}-\sum_{t=1}^6 \frac{y^{(1)}_i(t)}{h^{(1)}_{i3}(t)}=L_{i\backslash3}(\xu^{[1|1]})+L_{i\backslash3}(\xu^{[2|1]}), \label{Eq:PSIN_Output3} 
\end{align}
where we denoted the transmitted vector of $S_k$ by $\xu^{[k|1]}\Def[u^{[k|1]}_1,\cdots,u^{[k|1]}_6]^T$ and its contribution in the received signal of $V^{(2)}_i$ by $L_{i\backslash j}(\xu^{[k|1]})$ after nulling the effect of $S_{j}$. More specifically,
\begin{align}
L_{i \backslash j}(\xu^{[k|1]}) \Def \sum_{t=1}^6\left(\frac{h^{(1)}_{ik}(7)}{h^{(1)}_{ij}(7)}{-}\frac{h^{(1)}_{ik}(t)}{h^{(1)}_{ij}(t)}\right)u^{[k|1]}_t, \hspace{1cm} 1\leq i,j,k \leq 3,\hspace{5mm} j\neq k.
\end{align}

Next, we show that the six linear combinations $\{L_{i \backslash j}(\xu^{[k|1]})\}_{i=1}^3$, $j\in \{1,2,3\} \backslash \{k\}$, are linearly independent almost surely. We show this only for $k=1$ since the other cases are similar. The coefficient matrix of these six linear combinations is as follows.
\begin{align}
\xC_{6\times 6} = \xA_{6\times 6}-\xB_{6\times 6},
\end{align}
where
\begin{align}
\xA \Def \diag\left( a_1,a_2,\cdots,a_6 \right) \cdot [ 1 ]_{6 \times 6}\hspace{5mm}\textup{with}\hspace{5mm} 
a_\ell=\begin{cases}
\frac{h^{(1)}_{\ell1}(7)}{h^{(1)}_{\ell2}(7)}, & 1\leq \ell \leq 3 \\
\frac{h^{(1)}_{(\ell-3)1}(7)}{h^{(1)}_{(\ell-3)3}(7)}, & 4 \leq \ell \leq 6
\end{cases},
\end{align}
and
\begin{align}
\xB \Def [ b_{ij}]_{6\times 6} \hspace{5mm}\textup{with}\hspace{5mm} 
b_{ij}=
\begin{cases}
\frac{h^{(1)}_{i1}(j)}{h^{(1)}_{i2}(j)}, & 1\leq i \leq 3 \\
\frac{h^{(1)}_{(i-3)1}(j)}{h^{(1)}_{(i-3)3}(j)}, & 4 \leq i \leq 6
\end{cases}.
\end{align}
Since the channel coefficients are i.i.d. over time and nodes, $b_{ij}$'s are independent of $a_\ell$'s. Moreover, denominators of $b_{ij}$'s are independent of their nominators. Thus, it can be easily seen that, given the \emph{non-zero} channel coefficients $\{h^{(1)}_{ij}(7) : 1\leq i,j \leq 3\}$ and $\{h^{(1)}_{i1}(j) : 1\leq i\leq 3 , 1\leq j \leq 6\}$, the $36$ elements of matrix $\xC$ are independent of each other (with continuous conditional distributions), and hence, it is full rank. Therefore, since the probability of a channel coefficient being zero is zero, $\xC$ is full rank almost surely.

Thus, if all of these six linear combinations are delivered to $D_{1}$, it will be able to solve them for $\xu^{[k|1]}$. The rest of the transmission scheme is dedicated to this goal. Equivalently, for an arbitrary $N_1$, $N_1$ symbols can be transmitted in $T^{(1)}_1=7\times \frac{N_1}{18}$ time slots in hop $1$ and $9\times \frac{N_1}{18}=\frac{N_1}{2}$ linear combinations are generated at $\{V^{(2)}_i\}_{i=1}^3$ as in \cref{Eq:PSIN_Output1,Eq:PSIN_Output2,Eq:PSIN_Output3}.

\hop{$2$ (AF)}

AF operation is performed by $\{V^{(2)}_i\}_{i=1}^3$ in hop $2$. In particular, $\{V^{(2)}_i\}_{i=1}^3$ amplify and forward the linear combinations obtained in hop $1$, \ie \cref{Eq:PSIN_Output1,Eq:PSIN_Output2,Eq:PSIN_Output3}, over hop $2$ at three linear combinations per time slot. For instance, in one time slot, $\{L_{i\backslash1}(\xu^{[2|1]})+L_{i\backslash1}(\xu^{[3|1]})\}_{i=1}^3$ are transmitted respectively by $\{V^{(2)}_i\}_{i=1}^3$ and
\begin{align}
\{L'_{j\backslash1}(\xu^{[2|1]})+L'_{j\backslash1}(\xu^{[3|1]})\}_{j=1}^3 \label{Eq:3user3hop-AF-phase1_1}
\end{align}
are received respectively by $\{V^{(3)}_j\}_{j=1}^3$, where 
\begin{align}
L'_{j\backslash 1}(\xu^{[k|1]})\Def\sum_{i=1}^3 h_{ji}^{(2)}(t)L_{i\backslash 1}(\xu^{[k|1]}). \label{Eq:3user3hop-AF-phase1_2}
\end{align}
Hence, this hop takes $T^{(2)}_1=\frac{N_1}{2} \times \frac{1}{3}=\frac{N_1}{6}$ time slots.

\hop{$3$ (Generation of Order-$2$ Symbols)}

Relays $\{V^{(3)}_j\}_{j=1}^3$ amplify and forward the signals received during hop $2$, cf. \cref{Eq:3user3hop-AF-phase1_1,Eq:3user3hop-AF-phase1_2}. Recall that if all the $L_{i\backslash j'}(\xu^{[k|1]})$'s or, equivalently, all the $L'_{i\backslash j'}(\xu^{[k|1]})$'s are delivered to $D_1$, then it will be able to decode all of its information symbols. Let $\{V^{(3)}_j\}_{j=1}^3$ spend one time slot to transmit $\{L'_{j\backslash1}(\xu^{[2|1]})+L'_{j\backslash1}(\xu^{[3|1]})\}_{j=1}^3$, respectively, which include six quantities of type $L'$. Then, $\{L''_{j\backslash1}(\xu^{[2|1]})+L''_{j\backslash1}(\xu^{[3|1]})\}_{j=1}^3$ are received respectively by $\{D_j\}_{j=1}^3$, where 
\begin{align}
L''_{j\backslash 1}(\xu^{[k|1]})\Def\sum_{i=1}^3 h_{ji}^{(3)}(t)L'_{i\backslash 1}(\xu^{[k|1]}),\hspace{5mm} 1\leq j \leq 3. \label{Eq:O2Generation_3user}
\end{align}
We note that $D_1$ receives an entirely desired linear combination, \ie $L''_{1\backslash1}(\xu^{[2|1]})+L''_{1\backslash1}(\xu^{[3|1]})$. Moreover, if we deliver $L''_{2\backslash1}(\xu^{[2|1]})$ to both $D_1$ and $D_2$, then $D_1$ obtains another desired linear combination, whereas $D_2$ can cancel it out to obtain $L''_{2\backslash1}(\xu^{[3|1]})$, which, in turn, is desired by $D_1$. Hence, following \cref{Def:Order_m}, since $L''_{2\backslash1}(\xu^{[2|1]})$ can be reconstructed by $S_2$ using delayed CSI, we consider it as an order-$2$ symbol and denote it by $u^{[2|1,2]}$. Then, $L''_{2\backslash1}(\xu^{[3|1]})$ will be a side information available at $D_2$ and desired by $D_1$, which can be reconstructed by $S_3$ using delayed CSI. As such, we denote it by $u^{[3|1;2]}$. Similarly, the order-$2$ symbol $u^{[2|1,3]}\Def L''_{3\backslash1}(\xu^{[2|1]})$ and side information $u^{[3|1;3]}\Def L''_{3\backslash1}(\xu^{[3|1]})$ are generated at $D_3$. 

After delivering $u^{[2|1,2]}$, $u^{[3|1;2]}$, $u^{[2|1,3]}$, and $u^{[3|1;3]}$ to $D_1$, it will obtain five desired equations (including its own received one) in terms of the six transmitted quantities of type $L'$. Hence, it still needs another equation in terms of the transmitted $L'$ quantities to be able to decode all of them. To provide $D_1$ with the desired equation, the above time slot is repeated $6/5$ times. Equivalently, five time slots are spent similarly by transmitting five \emph{distinct} sets of $L'$ quantities. Then, another time slot is spent by transmitting summation of all of the five previously transmitted signals by each relay node.

In total, $T^{(3)}_1=\frac{N_1}{6}\times \frac{6}{5} = \frac{N_1}{5}$ time slots are spent in hop $3$ during phase $1$ and $\frac{N_1}{5}\times 2$ order-$2$ symbols together with $\frac{N_1}{5}\times 2$ side information quantities are generated. Finally, since \emph{each} source node has information symbols for \emph{each} destination, it can be shown that, by scheduling all of the destination nodes in this phase, an equal number of side information quantities of both types $u^{[i|j;k]}$ and $u^{[i|k;j]}$ are generated for any $1\leq i , j , k \leq 3$. Then, it is readily seen that for any  $1\leq i , j , k \leq 3$, the following quantity is available at $S_i$ and desired by both $D_j$ and $D_k$ and, thus, is a new order-$2$ symbol.
\begin{align}
u^{[i|j,k]}\Def u^{[i|j;k]} + u^{[i|k;j]}. \label{Eq:O2Generation_3user_1}
\end{align}
Hence, the generated side information symbols can be appropriately grouped into distinct pairs, such that each pair yields a new order-$2$ symbol. Therefore, the total number of generated order-$2$ symbols by the end of phase $1$ is $N_2=\frac{2N_1}{5}+\frac{N_1}{5}=\frac{3N_1}{5}$.

\phase{$2$}

The goal of phase $2$ is to transmit the order-$2$ symbols generated by the end of phase $1$ over the network. To this end, we distribute the transmission load over multiple hops by offloading the order-$2$ symbols from the source nodes to the relay nodes. Therefore, after offloading the symbols, the relays will be responsible for delivering the order-$2$ symbols to the destination nodes without further involvement of the source nodes. An important observation here is that, instead of offloading the original order-$2$ symbols to the relays, which is equivalent to transmission over a single-hop X channel with delayed CSIT, linearly transformed versions of them are delivered as \emph{new} order-$2$ symbols. This is beneficial in terms of achievable DoF.

\hop{$1$ (Symbol Offloading)}

The order-$2$ symbols are offloaded from the source nodes to $\{V^{(2)}_k\}_{k=1}^3$. Each time slot of this hop is dedicated to a pair of destination nodes. During the time slot dedicated to ($D_i$, $D_j$), $u^{[1|i,j]}$, $u^{[2|i,j]}$, and $u^{[3|i,j]}$ are transmitted by $S_1$, $S_2$, and $S_3$, respectively. During this time slot, relay $V^{(2)}_k$, $1\leq k \leq 3$, receives linear combination $y^{(1)}_k(t)=\sum_{\ell=1}^3 h^{(1)}_{k\ell}(t)u^{[\ell|i,j]}$ of the three transmitted order-$2$ symbols, where $t$ is the corresponding time slot. If all of these three linear combinations are delivered eventually to both $D_i$ and $D_j$, then both nodes will be able to decode $u^{[1|i,j]}$, $u^{[2|i,j]}$, and $u^{[3|i,j]}$. Therefore, $y^{(1)}_1(t)$, $y^{(1)}_2(t)$, and $y^{(1)}_3(t)$ can be considered as three \emph{new} order-$2$ symbols that are now available at the relay side (not the source side). Hence, $N_2$ new order-$2$ symbols are generated at layer-$2$ relays in this hop during $T^{(1)}_2=\frac{N_2}{3}=\frac{N_1}{5}$ time slots.

\hop{$2$ (PSIN)}

The same PSIN operation proposed for hop $1$ in phase $1$ is performed for transmission of (new) order-$2$ symbols over hop $2$ in phase $2$, and thus, $T^{(2)}_2=\frac{7N_2}{18}=\frac{7N_1}{30}$. In particular, $18$ ``new'' order-$2$ symbols, all desired by a specific pair of destinations, say ($D_1$, $D_2$), are transmitted by relays $\{V^{(2)}_i\}_{i=1}^3$ in seven time slots in the same way as the information symbols were transmitted by the source nodes over hop $1$ in phase $1$ (cf. \cref{Eq:3user3hop-PSIN-phase1_1,Eq:3user3hop-PSIN-phase1_2}). Then, after partial interference nulling, relay $V^{(3)}_i$, $1\leq i \leq 3$, obtains the linear combinations
\begin{align}
&L_{i\backslash1}(\xu^{[2|1,2]})+L_{i\backslash1}(\xu^{[3|1,2]}), \label{Eq:3user3hop-PSIN_phase2_1} \\
&L_{i\backslash2}(\xu^{[3|1,2]})+L_{i\backslash2}(\xu^{[1|1,2]}),\label{Eq:3user3hop-PSIN_phase2_2}  \\
&L_{i\backslash3}(\xu^{[1|1,2]})+L_{i\backslash3}(\xu^{[2|1,2]}), \label{Eq:3user3hop-PSIN_phase2_3} 
\end{align}
where we denoted the transmitted vector of $V^{(2)}_j$ by $\xu^{[j|1,2]}\Def[u^{[j|1,2]}_1,\cdots,u^{[j|1,2]}_6]^T$ and its contribution in the received signal of $V^{(3)}_i$ by $L_{i\backslash j'}(\xu^{[j|1,2]})$ after nulling the effect of $V^{(2)}_{j'}$, $j'\neq j$. Note that here, with a slight abuse of notation, the symbols $u^{[j|1,2]}$ denote the ``new'' order-$2$ symbols rather than the original ones.

\hop{$3$ (Generation of Order-$3$ Symbols)}

Relays $\{V^{(3)}_j\}_{j=1}^3$ amplify-and-forward the linear combinations obtained during hop $2$, cf. \cref{Eq:3user3hop-PSIN_phase2_1,Eq:3user3hop-PSIN_phase2_2,Eq:3user3hop-PSIN_phase2_3}. Recall that if all of the quantities $L_{i\backslash j'}(\xu^{[k|1,2]})$ are delivered to $D_1$ and $D_2$, then both of them will be able to decode all of their order-$2$ symbols. Let $\{V^{(3)}_j\}_{j=1}^3$ spend one time slot to transmit $\{L_{j\backslash1}(\xu^{[2|1,2]})+L_{j\backslash1}(\xu^{[3|1,2]})\}_{j=1}^3$, respectively. Then, $\{L'_{j\backslash1}(\xu^{[2|1,2]})+L'_{j\backslash1}(\xu^{[3|1,2]})\}_{j=1}^3$ are received, respectively, by $\{D_j\}_{j=1}^3$, where 
\begin{align}
L'_{j\backslash 1}(\xu^{[k|1,2]})\Def\sum_{i=1}^3 h_{ji}^{(3)}(t)L_{i\backslash 1}(\xu^{[k|1,2]}),\hspace{1cm} 1\leq j \leq 3,\hspace{1cm} k=2,3. \label{Eq:O3Generation_3user}
\end{align}
We observe that $D_1$ and $D_2$ each receive an entirely desired linear combination. Moreover, if we deliver $L'_{3\backslash1}(\xu^{[2|1,2]})$ to all three destinations, then $D_1$ and $D_2$ each obtain another desired linear combination whereas $D_3$ can cancel it out to obtain $L'_{3\backslash1}(\xu^{[3|1,2]})$, which in turn is desired by both $D_1$ and $D_2$. Hence, since $L'_{3\backslash1}(\xu^{[2|1,2]})$ can be reconstructed by $V^{(2)}_2$ using delayed CSI, we consider it as an order-$3$ symbol and denote it by $u^{[2|1,2,3]}$. Then, since $L'_{3\backslash1}(\xu^{[3|1,2]})$ will be available at $D_3$ and desired by both $D_1$ and $D_2$, and can be reconstructed by $V^{(2)}_3$ using delayed CSI, we denote it by $u^{[3|1,2;3]}$. 

After delivering $u^{[2|1,2,3]}$, $u^{[3|1,2;3]}$ to both $D_1$ and $D_2$, each of them obtains three desired equations (including its own received one) in terms of the six transmitted quantities of type $L$. Hence, each of them still requires three more equations in terms of the transmitted $L$ quantities. To provide them with the desired equations, the above time slot is repeated one more time.

In total, $T^{(3)}_2=\frac{N_2}{6}\times 2=\frac{N_2}{3}=\frac{N_1}{5}$ time slots are spent in hop $3$ during phase $2$ and $\frac{N_1}{5}$ order-$3$ symbols together with $\frac{N_1}{5}$ side information quantities are generated. Finally, since \emph{each} layer-$2$ relay node has order-$2$ symbols for \emph{each} pair of destinations, one can easily show that, by scheduling all pairs of destination nodes in this phase, an equal number of side information quantities of types $u^{[i|1,2;3]}$, $u^{[i|2,3;1]}$, and $u^{[i|3,1;2]}$ are generated for any $1\leq i \leq 3$. Note that each destination node wishes to obtain two of these three quantities and has the third one. Therefore, two random linear combinations of these three quantities are desired by all three destinations and, thus, are considered as two new order-$3$ symbols available at $V^{(2)}_i$. Hence, the generated side information symbols can be appropriately grouped into distinct triples, such that each triple yields two new order-$3$ symbols. Therefore, the total number of order-$2$ symbols generated by the end of phase $2$ is $N_3=\frac{N_1}{5}+\frac{N_1}{5}\times \frac{2}{3}=\frac{N_1}{3}$.

\phase{$3$}

The goal of this phase is to deliver the order-$3$ symbols, generated by the end of phase $2$, to all destinations. As a result of symbol offloading in the previous phase, hop $1$ is silent in this phase. Hence, this phase starts with transmission over hop $2$.

\hop{$2$ (Symbol Offloading)}

Similar to hop $1$ in phase $2$, the order-$3$ symbols are offloaded by $\{V^{2)}_j\}_{j=1}^3$ to $\{V^{3)}_j\}_{j=1}^3$ at three symbols per time slot. Therefore, the total duration of this hop in phase $3$ is $T^{(2)}_3=\frac{N_3}{3}=\frac{N_1}{9}$.

\hop{$3$ (Final Delivery)}

The offloaded order-$3$ symbols are delivered by $\{V^{3)}_j\}_{j=1}^3$ at one symbol per time slot to all destinations using a time division scheme. Therefore, $T^{(3)}_3=N_3=\frac{N_1}{3}$.

\begin{figure}
\centering
\includegraphics[width=\textwidth]{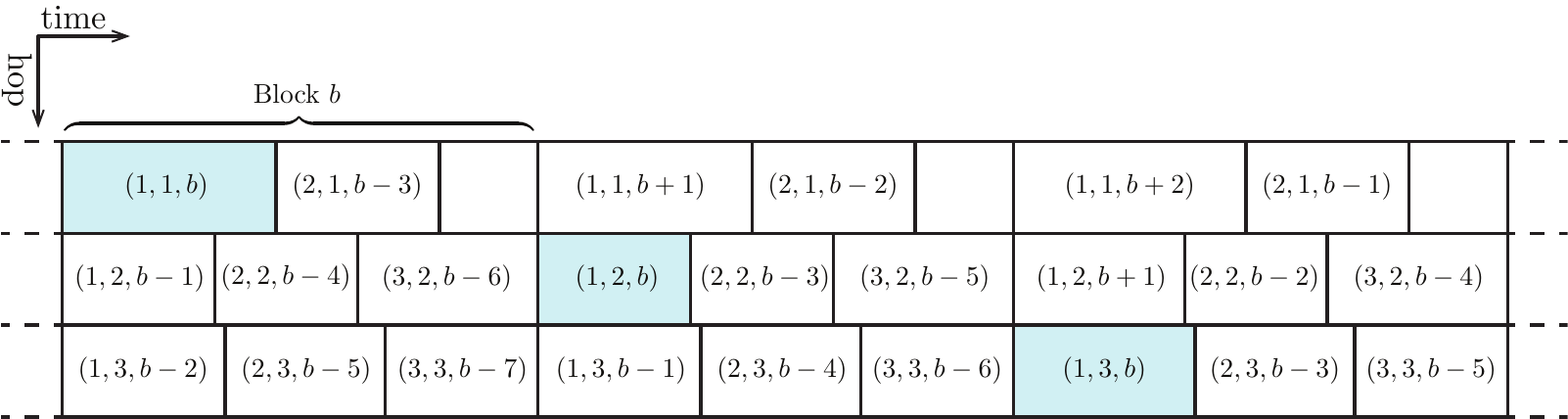}
\caption{Phase-hop interleaver for the $3$-user $3$-hop X network: The sub-block $(m,k,b)$ denotes transmission in hop $k$ during phase $m$ in round $b$. For any $1\leq b \leq B$, the sub-blocks $(1,1,b)$, $(1,2,b)$, $(1,3,b)$, $(2,1,b)$, $(2,2,b)$, $(2,3,b)$, $(3,2,b)$, and $(3,3,b)$ are accomplished in eight consecutive blocks.}
\label{Fig:3User3hop_interleaver}
\end{figure}

Finally, in order to achieve $15/11$ DoF, we perform $B$ rounds of the transmission scheme consecutively. The phases/hops of different rounds are interleaved such that $N_1B$ information symbols are transmitted in $B+7$ blocks, as depicted in \cref{Fig:3User3hop_interleaver}. The sub-block $(m,k,b)$ in the figure denotes transmission in hop $k$ during phase $m$ in round $b$. For any $1\leq b \leq B$, the sub-blocks $(1,1,b)$, $(1,2,b)$, $(1,3,b)$, $(2,1,b)$, $(2,2,b)$, $(2,3,b)$, $(3,2,b)$, and $(3,3,b)$ are accomplished in blocks $b, b+1, \cdots, b+7$, respectively, as shown in the figure. The time duration of each block in the interleaved scheme is $\max\{T^{(1)},T^{(2)},T^{(3)}\}$, where $T^{(k)}\Def T^{(k)}_1+T^{(k)}_2+T^{(k)}_3$ is the total time duration of hop $k$. Hence, the achieved DoF is equal to
\begin{align}
\lim_{B\to \infty}\frac{N_1B}{\max\{T^{(1)},T^{(2)},T^{(3)}\}(B+7)}=\frac{N_1}{\max\{T^{(1)},T^{(2)},T^{(3)}\}}, \label{Eq:DoF_3user3hop}
\end{align}
where
\begin{align}
T^{(1)}=T^{(1)}_1+T^{(1)}_2+T^{(1)}_3&= \frac{7N_1}{18}+\frac{N_1}{5}+0=\frac{53N_1}{90}, \nonumber \\
T^{(2)}=T^{(2)}_1+T^{(2)}_2+T^{(2)}_3&=\frac{N_1}{6}+\frac{7N_1}{30}+\frac{N_1}{9}=\frac{23N_1}{45}, \nonumber \\
T^{(3)}=T^{(3)}_1+T^{(3)}_2+T^{(3)}_3&=\frac{N_1}{5}+\frac{N_1}{5}+\frac{N_1}{3}=\frac{11N_1}{15}. \nonumber 
\end{align}
Therefore, the proposed scheme achieves $\frac{15}{11}$ DoF for the $3$-user $3$-hop X network.

\subsection{$K$-user $K$-hop X Network}
\label{Sec:KUser_KHop_Details}

The notations that are used throughout this section are provided below.
\begin{notation} 
The following notations are used in this section.
\begin{itemize}
\item[] $\xxI_m$: A subset of cardinality $m$ of $ \{1,2,\cdots,K\}$. Obviously, $\xxI_K= \{1,2,\cdots,K\}$
\item[] $\xxD(\xxI_m)$: Set of $m$ destination nodes that correspond to index set $\xxI_m$
\item[] $T^{(k)}_m$: Time duration of hop $k$ in phase $m$
\item[] $T^{(k)}$: Total time duration of hop $k$
\item[] $N_m$: Number of order-$m$ symbols that are transmitted during phase $m$ over the network
\end{itemize}
\end{notation}

Recall from \cref{Tbl:Operations} that the transmission strategy has $K$ phases. Phase $1$ of the scheme begins with transmission of the information symbols by the source nodes in hop $1$ and continues through subsequent hops up to hop $K$, wherein order-$2$ symbols are generated. The order-$2$ symbols are then transmitted during phase $2$. The transmission continues hop by hop in each phase, and phase by phase up to phase $K$, in which order-$K$ symbols are delivered to all destination nodes. Each phase involves transmission over a subset of hops from a specific hop on to hop $K$ and each involved hop is responsible for a specific operation, as indicated in \cref{Tbl:Operations}. These operations constitute the main building blocks of the scheme, and they are described separately in more detail in the following.

\subsubsection{PSIN in Hop $m$ of Phase $m$, $1\leq m \leq K-1$}
\label{Sec:PSIN_Details}

This is the first operation of the transmission scheme, which is performed in hop $1$ during phase $1$ and also in hop $m$ of phase $m$. Fix an integer $3\leq L\leq K$ throughout the scheme. Fix a subset of $L$ indices in $\xxI_K$ that, without loss of generality, is assumed to be $\{1,2,\cdots,L\}$, and also fix a subset $\xxI_m\subseteq \xxI_K$. In this hop, $KL(L-1)$ order-$m$ symbols, all desired by $\xxD(\xxI_m)$, are transmitted in $K(L-1)+1$ time slots as follows.

Each of $V^{(m)}_{1}$, $V^{(m)}_{2}$, $\cdots$, $V^{(m)}_{L}$ transmits $K(L-1)+1$ random linear combinations of $K(L-1)$ order-$m$ symbols, all desired by $\xxD(\xxI_m)$. Denote the vector of order-$m$ symbols transmitted by $V^{(m)}_{\ell}$ by
\begin{align}
\xu^{[\ell |\xxI_m]}\Def[u^{[\ell |\xxI_m]}_1,u^{[\ell |\xxI_m]}_2,\cdots,u^{[\ell |\xxI_m]}_{K(L-1)}]^T, \quad 1\leq \ell \leq L,
\end{align}
and denote the random precoding vector of relay $V^{(m)}_{\ell}$ in time slot $t$ by
\begin{align}
\xc^{[\ell |\xxI_m]}(t)\Def[c^{[\ell | \xxI_m]}_1(t),c^{[\ell |\xxI_m]}_2(t),\cdots,c^{[\ell | \xxI_m]}_{K(L-1)}(t)]^T,\quad 1\leq \ell \leq L,\quad 1\leq t \leq K(L-1)+1.
\end{align}
Then, ignoring the noise, the vector of signals that are received by relay $V^{(m+1)}_i$ during these time slots can be written as
\begin{align}
\xy^{(m)}_{i}=\xH^{(m)}_{i1}\xC^{[1|\xxI_m]}\xu^{[1|\xxI_m]}+\xH^{(m)}_{i2}\xC^{[2|\xxI_m]}\xu^{[2|\xxI_m]}+\cdots+\xH^{(m)}_{iL}\xC^{[L|\xxI_m]}\xu^{[L|\xxI_m]},\quad 1\leq i \leq K,
\end{align}
where
\begin{align}
\xH^{(m)}_{i\ell}&\Def \diag\left(h^{(m)}_{i\ell}(1),\cdots,h^{(m)}_{i\ell}(K(L-1)+1)\right),\quad 1\leq \ell \leq L, \\
\xC^{[\ell |\xxI_m]}&\Def\left[\xc^{[\ell |\xxI_m]}(1),\xc^{[\ell |\xxI_m]}(2),\cdots,\xc^{[\ell |\xxI_m]}(K(L-1)+1)\right]^T,\quad 1\leq \ell \leq L.
\end{align}
Since $\xH^{(m)}_{i\ell}\xC^{[\ell |\xxI_m]}$ is a $[K(L-1)+1]$-by-$[K(L-1)]$ matrix (which can be shown to be full rank almost surely), its left null space is one dimensional and is denoted by the vector $\xomega^{(m)}_{i\ell}$. Hence, $V^{(m+1)}_i$ can null out the effect of $\xu^{[\ell|\xxI_m]}$ from its received signals for any $1\leq \ell \leq L$ and obtain $L$ linear combinations
\begin{align}
\xxL^{(m)}_m(i\backslash \ell) \Def (\xomega^{(m)}_{i\ell})^T\xy^{(m)}_{i}= \sum_{\substack{\ell'=1\\ \ell'\neq \ell}}^L (\xomega^{(m)}_{i\ell})^T\xH^{(m)}_{i\ell'}\xC^{[\ell' |\xxI_m]}\xu^{[\ell'|\xxI_m]}= \sum_{\substack{\ell'=1\\ \ell'\neq \ell}}^L v_{i\backslash \ell}^{[\ell'|\xxI_m]},\quad 1\leq \ell \leq L, \label{Eq:PhmHm_LC}
\end{align}
where 
\begin{align}
v_{i\backslash \ell}^{[\ell'|\xxI_m]}\Def (\xomega^{(m)}_{i\ell})^T\xH^{(m)}_{i\ell'}\xC^{[\ell' |\xxI_m]}\xu^{[\ell'|\xxI_m]} \label{Eq:v_quantities}
\end{align}
is the partial linear combination (PLC) containing the entire contribution of $\xu^{[\ell'|\xxI_m]}$ in the received signal of $V^{(m+1)}_i$ after nulling $\xu^{[\ell|\xxI_m]}$. One can verify that for any $\ell$, the vector $\xu^{[\ell|\xxI_m]}$ contributes to all linear combinations $\xxL^{(m)}_m(i\backslash \ell')$ with $\ell' \neq \ell$, which yields a total of $K(L-1)$ PLCs, namely, $\{v_{i\backslash \ell'}^{[\ell|\xxI_m]}\}_{i=1}^K$, $\ell' \in \{1,\cdots, L\} \backslash \{\ell\}$. Moreover, it can be shown that these $K(L-1)$ contributions are indeed $K(L-1)$ \emph{linearly-independent} combinations of the elements of $\xu^{[\ell|\xxI_m]}$. Therefore, if they are finally delivered to $\xxD(\xxI_m)$, each of these destination nodes can decode all $K(L-1)$ order-$m$ symbols contained in $\xu^{[\ell|\xxI_m]}$.

In summary, in hop $m$ of phase $m$, $KL(L-1)$ order-$m$ symbols are transmitted by $\{V^{(m)}_\ell\}_{\ell=1}^L$ during $K(L-1)+1$ time slots, and $KL$ linear combinations $\{\xxL^{(m)}_m(i\backslash \ell)\}_{i=1}^K$, $1\leq \ell \leq L$, are obtained by relays $\{V^{(m+1)}_i\}_{i=1}^K$, as defined in \cref{Eq:PhmHm_LC}. Since $N_m$ order-$m$ symbols are transmitted in phase $m$, the number of spent time slots of hop $m$ in phase $m$ is equal to
\begin{align}
T_m^{(m)}=N_m\times\frac{K(L-1)+1}{KL(L-1)}, \quad 1\leq m \leq K-1. \label{Eq:T_m^(m)}
\end{align}

\subsubsection{Order-$m$ Symbol Offloading in Hop $m-1$ of Phase $m$, $2\leq m \leq K$}
\label{Sec:Offloading_Details}

This operation is performed in hop $m-1$ of phase $m$. In particular, in each time slot of this hop, each of the nodes $\{V^{(m-1)}_i\}_{i=1}^K$ transmits one order-$m$ symbol desired by $\xxD(\xxI_m)$ for a fixed subset $\xxI_m\subseteq \xxI_K$. Therefore, the number of spent time slots is given by
\begin{align}
T^{(m-1)}_m = \frac{N_m}{K}, \quad 2\leq m \leq K. \label{Eq:T_m^(m-1)}
\end{align}

\subsubsection{Amplify-and-Forward in Hops $m+1$ to $K-1$ of Phase $m$, $1\leq m \leq K-2$}
\label{Sec:AF_Details}

This operation is performed in hops $m+1$ to $K-1$ of phase $m$. Recall that for each set of $KL(L-1)$ order-$m$ symbols transmitted using PSIN in hop $m$ of phase $m$, $KL$ linear combinations $\{\xxL^{(m)}_m(i\backslash \ell)\}_{i=1}^K$, $1\leq \ell \leq L$, are obtained by relays $\{V^{(m+1)}_i\}_{i=1}^K$ (see \cref{Sec:PSIN_Details}). Accordingly, for each set of $KL(L-1)$ order-$m$ symbols transmitted in hop $m$ of phase $m$, $L$ time slots are spent in each of the hops $m+1$ to $K-1$ as follows. For any $m+1\leq k \leq K-1$, during the $\ell^{\textup{th}}$ time slot of hop $k$ in phase $m$, relays $\{V^{(k)}_i\}_{i=1}^K$ transmit $\{\xxL^{(k-1)}_m(i\backslash \ell)\}_{i=1}^K$, respectively. Consequently, the following $L$ linear combinations are received by each of the relays $\{V^{(k+1)}_i\}_{i=1}^K$.
\begin{align}
\xxL^{(k)}_m(i\backslash \ell) \Def y^{(k)}_i(t) = \sum_{j=1}^K h^{(k)}_{ij}(t) \xxL^{(k-1)}_m(j\backslash \ell), \quad 1\leq i \leq K, \quad 1\leq \ell \leq L.
\end{align}

Since $N_m$ order-$m$ symbols are transmitted in phase $m$, the number of spent time slots of hop $k$ in phase $m$ is equal to
\begin{align}
T_m^{(k)}=\frac{N_m}{KL(L-1)}\times L=\frac{N_m}{K(L-1)},\quad 1\leq m \leq K-2, \quad m+1\leq k \leq K-1. \label{Eq:T_m^(m+1_K-1)}
\end{align}

\subsubsection{Generation of Order-$(m+1)$ Symbols in Hop $K$ of Phase $m$, $1\leq m \leq K-1$}
\label{Sec:Order_m+1}

This operation is performed in hop $K$ of phase $m$. Assume that for each $1\leq \ell \leq L$, relays $\{V^{(K)}_i\}_{i=1}^K$ spend one time slot to amplify-and-forward $\{\xxL^{(K-1)}_m(i\backslash \ell)\}_{i=1}^K$. During this time slot, each destination node $D_i$ receives a linear combination of the transmitted quantities as follows.
\begin{align}
y^{(K)}_i(t) = \sum_{j=1}^K h^{(K)}_{ij}(t) \xxL^{(K-1)}_m(j\backslash \ell)&=\left(\xh_i^{(K)}(t)\right)^T \xH^{(K-1)}(t)\xH^{(K-2)}(t)\cdots\xH^{(m+1)}(t)\boldsymbol{\xxL}^{(m)}_{\ell} \nonumber\\
&=\left(\tilde{\xh}^{(K)}_{i,m}(t)\right)^T \boldsymbol{\xxL}^{(m)}_{\ell} \nonumber\\
& =\sum_{\substack{\ell' =1 \\ \ell' \neq \ell}}^L\left(\tilde{\xh}^{(K)}_{i,m}(t)\right)^T \xv_\ell^{[\ell'|\xxI_m]}, \quad 1\leq i \leq K, \label{Eq:Received_L-1_Part}
\end{align} 
where
\begin{align}
\boldsymbol{\xxL}^{(m)}_{\ell}&\Def \left[\xxL^{(m)}_m(1\backslash \ell),\xxL^{(m)}_m(2\backslash \ell),\cdots,\xxL^{(m)}_m(K\backslash \ell)\right]^T,\quad \quad 1\leq \ell \leq L \\
\left(\tilde{\xh}^{(K)}_{i,m}(t)\right)^T &\Def \left(\xh_i^{(K)}(t)\right)^T \xH^{(K-1)}(t)\xH^{(K-2)}(t)\cdots\xH^{(m+1)}(t), \quad 1\leq i \leq K,\\
\xv_\ell^{[\ell'|\xxI_m]} & \Def \left[v_{1\backslash \ell}^{[\ell'|\xxI_m]},v_{2\backslash \ell}^{[\ell'|\xxI_m]},\cdots,v_{K\backslash \ell}^{[\ell'|\xxI_m]}\right]^T,\quad \quad 1\leq \ell, \ell' \leq L,\quad \ell' \neq \ell.
\end{align}

We note that for any $1\leq \ell \leq L$, the $K(L-1)$ ``$v$ quantities'' defined in \cref{Eq:v_quantities}, \ie
\begin{align}
v_{1\backslash \ell}^{[\ell' |\xxI_m]},v_{2\backslash \ell}^{[\ell' |\xxI_m]},\cdots,v_{K\backslash \ell}^{[\ell' |\xxI_m]},\quad \ell' \in \{1,\cdots, L\} \backslash \{\ell\},
\end{align}
which are all desired by $\xxD(\xxI_m)$, are transmitted in hop $K$ during one time slot. Now, for a fixed $\ell$, we have the following observations: 
\begin{enumerate}[(a)]
\item \label{Item:Om_extraLC} During this time slot, each destination node in $\xxD(\xxI_m)$ receives an equation solely in terms of the (desired) $v$ quantities and, hence, requires extra $K(L-1)-1$ linearly-independent equations to decode all of the $K(L-1)$ transmitted $v$ quantities.
\item \label{Item:Om}The equation received by $D_i$, $i \in \xxI_K \backslash \xxI_m$, is composed of $L-1$ PLCs as indicated in \cref{Eq:Received_L-1_Part}. Each of these PLCs is desired by $\xxD(\xxI_m)$. This totals $(K-m)(L-1)$ desired linear combinations. Also, each of them can be regenerated by a node $V^{(m)}_{\ell'}$, $\ell' \in \{1,\cdots ,L\} \backslash \{\ell\}$, using delayed CSI.
\item \label{Item:Om_repetition} If we deliver the mentioned PLCs to $\xxD(\xxI_m)$, then each of these $m$ destination nodes obtains $(K-m)(L-1)+1$ linear combinations (including its own received equation) in terms of the $K(L-1)$ transmitted $v$ quantities. Thus, each of them will still need $K(L-1)-[(K-m)(L-1)+1]=m(L-1)-1$ linearly-independent equations to be able to decode all transmitted $v$ quantities. Therefore, to provide $\xxD(\xxI_m)$ with a sufficient number of equations, this time slot is repeated $\frac{K(L-1)}{(K-m)(L-1)+1}$ times. This potentially leads to a fractional number of time slots which can be overcome simply by appropriate repetition of phase $m$.

Therefore, this hop takes a total of
\begin{align}
T_m^{(K)}=T^{(K-1)}_m\times \frac{K(L-1)}{(K-m)(L-1)+1}= \frac{N_m}{(K-m)(L-1)+1}, \quad 1\leq m \leq K-1 \label{Eq:T_m^(K)}
\end{align}
time slots, where the second equality uses \cref{Eq:T_m^(m+1_K-1)}.

\item Following observation (\ref{Item:Om}), if we deliver $L-2$ out of the $L-1$ PLCs of $D_i$, $i \in \xxI_K \backslash \xxI_m$, to $D_i$, it can cancel them to obtain the last PLC. Since each of them is also desired by $\xxD(\xxI_m)$, they are considered as $L-2$ order-$(m+1)$ symbols desired by $\xxD(\xxI_m)\cup \{D_i\}$. This yields a total of $(K-m)(L-2)$ order-$(m+1)$ symbols per time slot. Moreover, the last PLC (called remaining PLC) of each $D_i$, $i \in \xxI_K \backslash \xxI_m$, is considered as $\frac{m}{m+1}$ of an order-$(m+1)$ symbol yielding another $\frac{m}{m+1}(K-m)$ order-$(m+1)$ symbols. This is due to the fact that we can repeat this phase with appropriate permutation of the users such that for each subset $\{i_1,i_2,\cdots,i_{m+1}\}\subseteq \{1,2,\cdots,K\}$, $m+1$ remaining PLCs exist, each of which
\begin{itemize}
\item is available at one of $D_{i_1}$, $D_{i_2}$, $\cdots$, and $D_{i_{m+1}}$ and is desired by the other $m$
\item can be generated by the same node out of $\{V_i^{(m)}\}_{i=1}^K$
\end{itemize}
Therefore, $m$ random linear combinations of these $m+1$ remaining PLCs form $m$ order-$(m+1)$ symbols, or equivalently, each of them is $\frac{m}{m+1}$ of an order-$(m+1)$ symbol.

Hence, the total number of order-$(m+1)$ symbols is given by
\begin{align}
N_{m+1}&=T^{(K)}_m\times\left((K-m)(L-2)+\frac{m}{m+1}(K-m)\right)\nonumber \\
&=N_m\times \frac{(K-m)\left((m+1)(L-1)-1\right)}{(m+1)\left((K-m)(L-1)+1\right)}, \hspace{2cm} 1 \leq m \leq K-1, \label{Eq:N_m+1}
\end{align}
where \cref{Eq:N_m+1} is obtained by using \cref{Eq:T_m^(K)}.
\end{enumerate}

\subsubsection{Silent Hops in Phase $m$, $3\leq m \leq K$}
\label{Sec:SilentHops}

Hops $1$ to $m-2$ are silent during phase $m$. Therefore,
\begin{align}
T_m^{(k)}=0,\quad\quad 1\leq k \leq m-2. \label{Eq:T_m^(1_m-2)}
\end{align}

\subsubsection{Final Delivery in Hop $K$ of Phase $K$}
\label{Sec:FinalDelivery}

Hop $K$ in phase $K$ is responsible for delivery of order-$K$ symbols to all $K$ destination nodes. Using a time division scheme, one order-$K$ symbol per time slot is delivered to the destination nodes, and thus, we have
\begin{align}
T^{(K)}_K=N_K. \label{Eq:T_K^(K)}
\end{align}

\section{Achievable DoF Analysis for the $K$-user Network}
\label{Sec:DoF_Analysis}

\begin{figure}
\centering
\includegraphics[width=\textwidth]{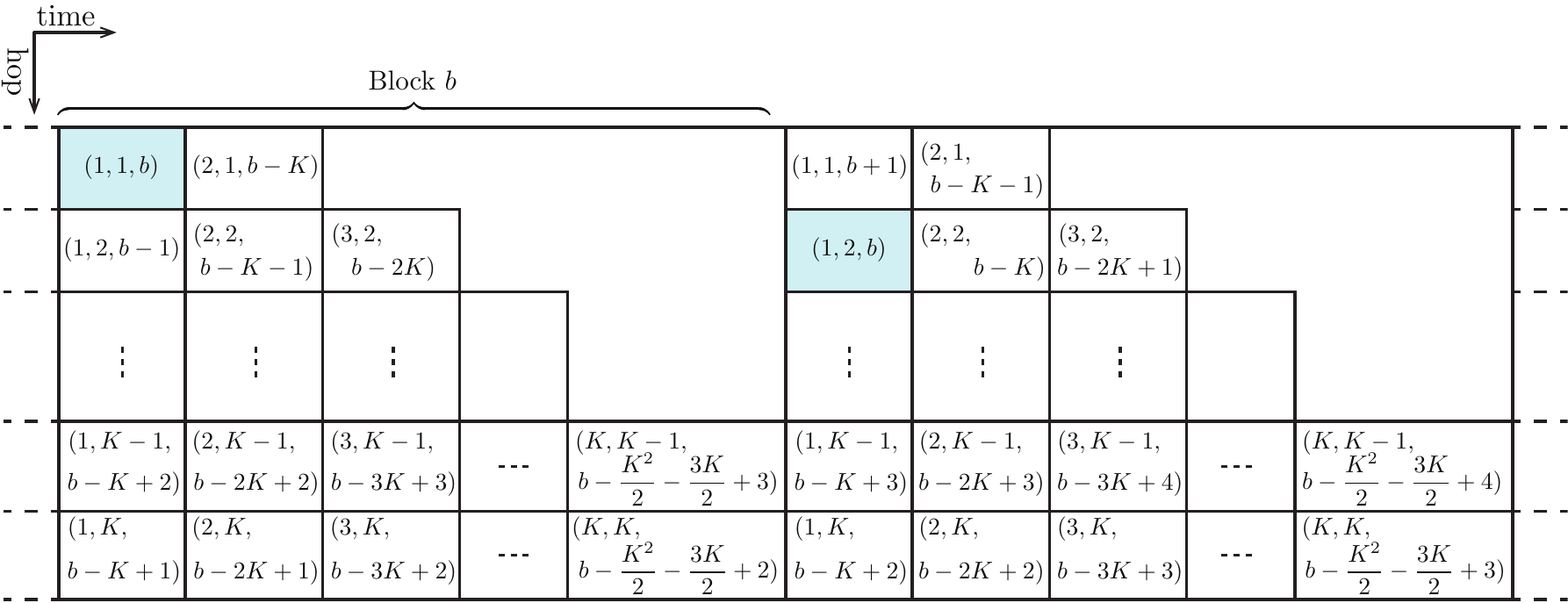}
\caption{Phase-hop interleaver for the $K$-user $K$-hop X channel: The sub-block $(m,k,b)$ denotes transmission in hop $k$ during phase $m$ in round $b$.}
\label{Fig:K_user_interleaver}
\end{figure}

In this section, we calculate the achievable DoF of the transmission scheme that was presented in \cref{Sec:Overview,Sec:AnalyticalDetails}. Similar to the $3$-user case, first, we apply a phase-hop interleaver to ensure that all hops are utilized effectively. More specifically, as illustrated in \cref{Fig:K_user_interleaver}, $BN_1$ information symbols are transmitted in $B$ rounds of the scheme over $B+\frac{K^2+3K}{2}-2$ consecutive blocks, each of which has duration $\max_{1\leq k \leq K} T^{(k)}$. Each block in \cref{Fig:K_user_interleaver} consists of $\frac{K^2+3K}{2}-1$ sub-blocks. For any $1\leq b \leq B$, the sub-block $(m,k,b)$ denotes the transmission in hop $k$ during phase $m$ in round $b$. For simplicity of illustration, the time durations of the sub-blocks are depicted to be the same in the figure. In fact, each sub block has a different duration, and unlike what is shown in the figure, $\max_{1\leq k \leq K} T^{(k)}$ is not determined by hop $K$ necessarily.
Using this interleaver, the achievable DoF is given by
\begin{align}
\DoF^\textup{X}(K)\geq \lim_{B\to \infty}\frac{B}{B+\frac{K^2+3K}{2}-2}\times \frac{N_1}{\max_{1\leq k \leq K} T^{(k)}}=\frac{N_1}{\max_{1\leq k \leq K} T^{(k)}}. \label{Eq:DoF_definition}
\end{align}

Now, we show that the achievable DoF of the proposed transmission scheme for the $K$-user $2K$-hop interference network is given by \cref{Th:2K_hop_IC_DoF}, which is repeated here for convenience.

\addtocounter{theorem}{-2} 

\begin{theorem}
The DoF of the $K$-user $2K$-hop interference network with $K\geq 3$ and delayed CSI satisfies
\begin{align}
\DoF^\textup{IC}(K,2K) \geq \frac{1}{t_1(q,K)+t_2(q,K)}, \label{Eq:DoF_lowerbound_repeated}
\end{align}
where
\begin{align}
t_1(q,K)&\Def \frac{1}{q-1}\left(\frac{\Gamma(q^{-1})(K-1)!}{\Gamma(K+q^{-1})}-\frac{1}{K}\right), \\
t_2(q,K)&\Def \frac{Kq+1}{q(q+1)K}+\frac{(2q-1)(K-1)}{2K[(K-1)q+1]},
\end{align}
and $2\leq q \leq K-1$ is an arbitrary integer, and $\Gamma(\cdot)$ is the gamma function.
\end{theorem}

As a consequence of \cref{Th:2K_hop_IC_DoF}, we showed in \cref{Cor:DoFScaling} that multi-hopping provides DoF scaling in the $K$-user interference network with delayed CSI.

\begin{IEEEproof}[Proof of \Cref{Th:2K_hop_IC_DoF}]
Using \cref{Eq:T_m^(K),Eq:T_K^(K)}, one can calculate the total time duration of hop $K$ as
\begin{align}
T^{(K)}=\sum_{i=1}^{K} T_i^{(K)}=\sum_{i=1}^{K}\frac{N_i}{(L-1)(K-i)+1}. \label{Eq:TK}
\end{align}

Also, using \cref{Eq:T_m^(1_m-2),Eq:T_m^(m-1),Eq:T_m^(m),Eq:T_m^(m+1_K-1)}, one can calculate the total time duration of hop $k$ as
\begin{align}
T^{(k)}=\sum_{i=1}^{K} T_i^{(k)}=\sum_{i=1}^{k-1}\frac{N_i}{K(L-1)}+\frac{N_k(K(L-1)+1)}{KL(L-1)} + \frac{N_{k+1}}{K},\quad\quad 1\leq k \leq K-1. \label{Eq:Tk}
\end{align}

In Appendix \ref{App:Inequality}, we prove the following inequality.
\begin{align}
T^{(k)} \leq T^{(1)}+T^{(K)}, \quad 2\leq k \leq K-1. \label{Eq:Relaxed_Ineq1}
\end{align}
Combining \cref{Eq:DoF_definition,Eq:Relaxed_Ineq1}, one can write
\begin{align}
\DoF^\textup{X}(K)\geq \frac{N_1}{T^{(1)}+T^{(K)}}. \label{Eq:DoF_ineq}
\end{align}
Defining the normalized time duration of hop $k$ as $\bar{T}^{(k)}\Def \frac{1}{N_1}T^{(k)}$, we have
\begin{align}
\DoF^\textup{X}(K)\geq \frac{1}{\bar{T}^{(1)}+\bar{T}^{(K)}}. \label{Eq:DoF_ineq1}
\end{align}
Defining $\alpha \Def \frac{1}{L-1}$, it is shown in Appendix \ref{App:TK_ClosedForm} that
\begin{align}
\bar{T}^{(K)}=\frac{\alpha}{1-\alpha}\left(\frac{\Gamma(\alpha)(K-1)!}{\Gamma(K+\alpha)}-\frac{1}{K}\right),\quad 0< \alpha<1. \label{Eq:TbarK}
\end{align}
Also, from \cref{Eq:Tk}, we have
\begin{align}
\bar{T}^{(1)}=\frac{\alpha(K+\alpha)}{(1+\alpha)K}+\frac{(2-\alpha)(K-1)}{2K(K+\alpha-1)}. \label{Eq:Tbar1}
\end{align}
Proof of \Cref{Th:2K_hop_IC_DoF} is complete in view of \cref{Eq:DoF_ineq1,Eq:TbarK,Eq:Tbar1} and by replacing $\alpha$, $\bar{T}^{(K)}$, and $\bar{T}^{(1)}$ with $\frac{1}{q}$, $t_1(q,K)$, and $t_2(q,K)$, respectively, in the theorem statement.
\end{IEEEproof}

\begin{figure}[t]
\centering
\includegraphics{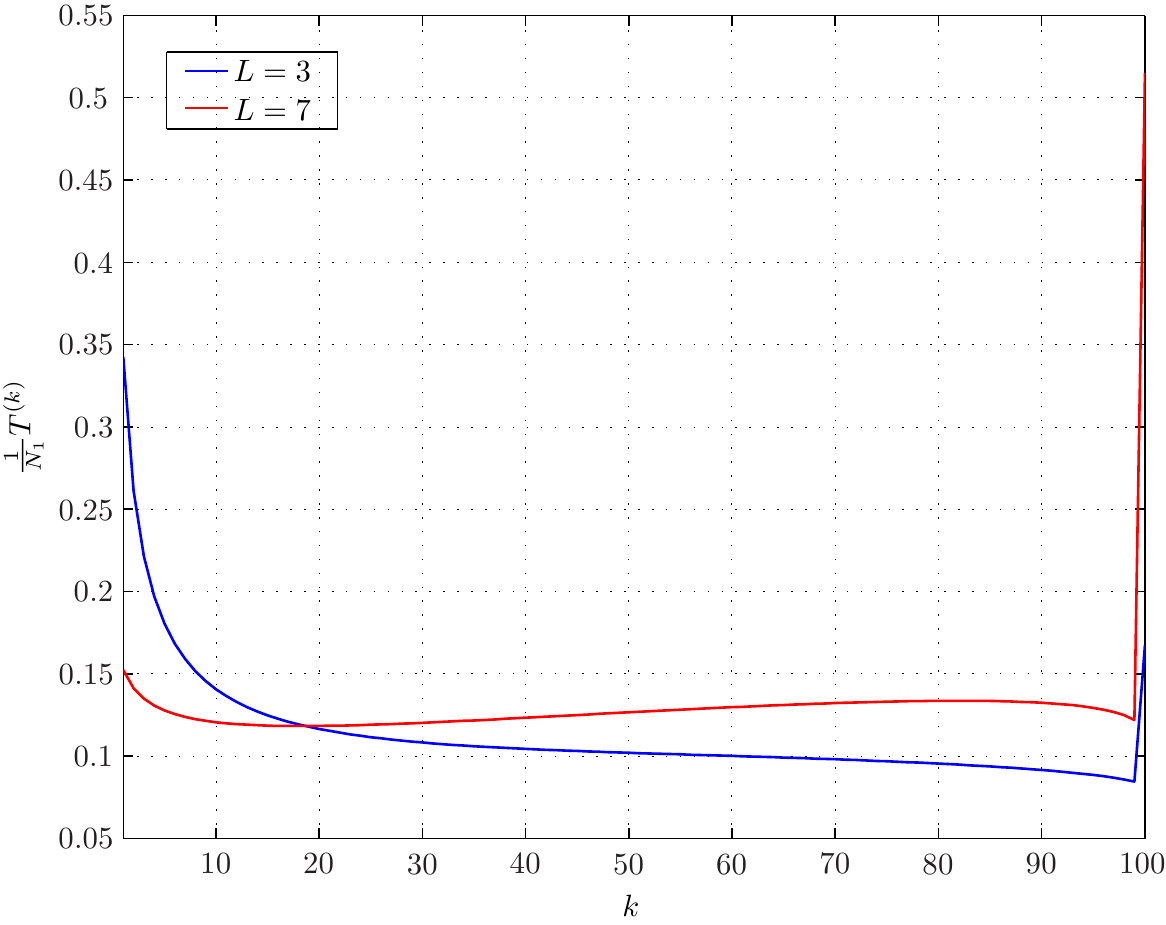}
\caption{Normalized time duration of hop $k$, \ie $\frac{1}{N_1}T^{(k)}$, in the $K$-user $K$-hop X network with $K=100$ for $L=3,7$.}
\label{Fig:NormalizedHopDuration}
\end{figure}

\begin{remark}
\label{Remark:NumericalDoF}
Numerical calculations demonstrate that $\bar{T}^{(k)} \leq \max\{\bar{T}^{(1)},\bar{T}^{(K)}\}$ for any $2\leq k \leq K-1$ and $3\leq L \leq K$. For instance, this can be seen in \cref{Fig:NormalizedHopDuration}, which shows $\bar{T}^{(k)}$ as a function of $k$ for $K=100$ and $L=3,7$. In other words, in view of \cref{Eq:DoF_definition}, the following lower bound can be verified numerically for any $K\geq 3$ and $2\leq q \leq K-1$.
\begin{align}
\DoF^\textup{X}(K)\geq \frac{1}{\max\{t_1(q,K),t_2(q,K)\}}. \label{Eq:ActualDoF}
\end{align}
Indeed, inequality \eqref{Eq:DoF_lowerbound_repeated} is a relaxed and more tractable version of inequality \cref{Eq:ActualDoF} and affects the scaling behaviour of the achievable DoF only with a factor of $1/2$.
\end{remark}

\begin{remark}
\label{Remark:DoFTable}
Our achievable DoF for the $K$-user $2K$-hop interference network with $K=3,5,10,20$ is listed in \Cref{Tbl:AchiecableDoF} and compared with the best known upper bound, \ie the $K$-user MISO broadcast channel upper bound \cite{maddah2012completely}. Note that this table presents our actual achievable DoF, given by \cref{Eq:ActualDoF} with $q=q^*(K)$, where
\begin{align}
q^*(K) \Def \argmax_{2\leq q \leq K-1} \frac{1}{\max \{t_1(q,K),t_2(q,K)\}}.
\end{align}
The table demonstrates how the gap between our achievable DoF and the best known upper bound increases with the number of users. In fact, it can be shown that this gap scales with the number of users, and thus, the problem of characterizing the DoF scaling rate of this network with delayed CSI remains open.
\end{remark}

\begin{table*}[t]
\caption{Our achievable DoF for the $K$-user $2K$-hop interference network with delayed CSI}
{\small
\begin{center}
\begin{tabular}{M{5cm}M{2cm}M{2cm}M{2.5cm}M{2.5cm}}
\toprule
$K$ & $3$ & $5$ & $10$ & $20$ \\
\midrule
\midrule
Achievable DoF for the $K$-user $2K$-hop interference network & $\frac{15}{11}\approx 1.364$ & $\frac{315}{193}\approx 1.632$ & $\frac{92378}{43191}\approx 2.139$ & $\frac{156}{59}\approx 2.644$ \\
\midrule
$K$-user MISO broadcast channel upper bound \cite{maddah2012completely} & $\frac{18}{11}\approx 1.636$ & $\frac{300}{137}\approx 2.190$ & $\frac{25200}{7381}\approx 3.414$ & $\frac{62078016}{11167027}\approx 5.559$\\
\bottomrule
\end{tabular}
\end{center}}
\label{Tbl:AchiecableDoF}
\end{table*}

\begin{remark}
\label{Remark:Mhop}
Although we proved the DoF lower bound of \cref{Th:2K_hop_IC_DoF} for the $K$-user $M$-hop interference network with delayed CSI and $M=2K$, it is indeed valid for any $M>2K$. To see this, consider a $K$-phase scheme in which
\begin{itemize}
\item hops $1, \cdots, 2K$ perform the same operations as in our proposed scheme, 
\item hops $2K+1,\cdots M$ perform AF operations.
 \end{itemize}
Then, the higher-order symbols are defined in the same way as before, now using the signals received by the destinations in hop $M$ (not hop $2K$). Since the AF operations in hops $2K+1,\cdots, M$ are accomplished at the maximum DoF of $K$ linear combinations per time slot, their durations are not greater than those of hops $1,\cdots, 2K$. Hence, we have $\max\{T^{(1)},\cdots, T^{(M)}\}=\max\{T^{(1)},\cdots,T^{(2K)}\}$. This implies that our achievable DoF of the $K$-user $2K$-hop network is also achievable in the $K$-user $M$-hop network with delayed CSI and $M>2K$.
\end{remark}

\begin{remark}
\label{Remark:Spread}
One round of the proposed scheme is spread over $\sum_{k=1}^{K}T^{(k)}$ time slots for transmission of $N_1$ information symbols. This is while the achievable DoF is governed by $\max_{1\leq k \leq K} T^{(k)}$ according to \cref{Eq:DoF_definition}.
\end{remark}

\section{$3$-user Multi-hop Interference Network: DoF Improvement}
\label{Sec:3_user_multi_hop_interference}
\begin{figure}[t]
\centering
\includegraphics{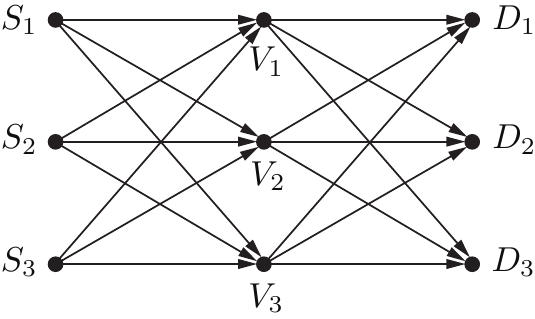}
\caption{$3$-user $2$-hop interference network.}
\label{Fig:3_user_2_hop_interference}
\end{figure}

In \cref{Sec:3user3hop}, we proved the achievability of $15/11$ DoF in the $3$-user $6$-hop interference network with delayed CSI. In this section, we show that this result is not tight. In particular, we prove the achievability of $16/11$ DoF in the $3$-user $2$-hop interference network with delayed CSI, as suggested by \cref{Th:3_user_2_hop_IC_DoF}. First, we show how we can achieve $36/25$ DoF in this network, depicted in \cref{Fig:3_user_2_hop_interference}, by proposing a $2$-phase transmission scheme. Then, we show the achievability of $16/11$ DoF by further optimizing the proposed scheme. Before proceeding with details of the scheme, let us highlight the main observations that are taken into account in the design of the improved scheme.

\begin{enumerate}
\item 
\label{Item:O2efficiency}
\emph{Order-$2$ efficiency of scheme}: Ideally, an order-$2$ symbol is a piece of information that, if delivered to its intended pair of destinations, provides \emph{each} of them with one desired linear combination in terms of its information symbols. However, this ideal condition may not be satisfied by all order-$2$ symbols in a transmission scheme. For example, consider the order-$2$ symbol $u^{[2|1,2]}=L''_{2\backslash1}(\xu^{[2|1]})$, which was defined in \cref{Eq:O2Generation_3user} at the end of phase $1$ of the scheme proposed in \cref{Sec:3user3hop}. This symbol provided $D_1$ with a desired linear combination after being delivered to both $D_1$ and $D_2$. However, $D_2$ used it to obtain $L''_{2\backslash1}(\xu^{[3|1]})$, which was another linear combination desired by $D_1$ (not $D_2$). In contrast, the order-$2$ symbols defined in \cref{Eq:O2Generation_3user_1} had the maximum efficiency, since each of them provided each of its intended destinations with a desired linear combination. To quantify this observation, we define \emph{order-$2$ efficiency} of a transmission scheme, denoted by $\eta_2$, as
\begin{align}
\eta_2\Def \frac{N_I}{2N_2},
\end{align}
where $N_2$ is the total number of order-$2$ symbols generated by the scheme, and $N_I$ is the total number of desired linear combinations that are provided for the destinations by the generated order-$2$ symbols. Clearly, $0\leq \eta_2 \leq 1$. It can be easily verified that for the scheme proposed in \cref{Sec:3user3hop}, we have 
\begin{align}
\eta_2=\frac{\frac{2N_1}{5}\times 1 + \frac{N_1}{5}\times 2}{2\times\frac{3N_1}{5}}\approx 0.67.
\end{align}

In order to utilize the available channel uses more efficiently and, hence, achieve higher DoF, transmission schemes with higher order-$2$ efficiency are desirable.

\item 
\label{Item:BalancedUtilization}
\emph{Balanced utilization of the hops}: It can be seen from  \cref{Eq:DoF_3user3hop} that the hop with maximal duration governs the achievable DoF of a transmission scheme. Therefore, all hops have the same duration in a DoF optimal scheme. Otherwise, there exists at least one underutilized hop, which can be fully utilized to yield higher achievable DoF.
\end{enumerate}

\subsection{$3$-user $2$-hop Interference Network: Achievability of $36/25$ DoF}
\label{Sec:3_user_2_hop_interference}

The transmission scheme is a $2$-phase scheme, which is outlined in \cref{Tbl:3User2hop_Operations} and described as follows.

\begin{table*}
\caption{Operations of different hops in the $2$-phase transmission scheme for the $3$-user $2$-hop interference network}
{\small
\begin{center}
\begin{tabular}{M{0.9cm}M{3cm}M{5.5cm}}
\toprule
Phase & Hop $1$ & Hop $2$\\
\midrule
\midrule
$1$ & PSIN & generation of order-$2$ symbols\\
\midrule
$2$ & symbol offloading & PSIN / generation of order-$3$ symbols / final delivery\\
\bottomrule
\end{tabular}
\end{center}}
\label{Tbl:3User2hop_Operations}
\end{table*}

\phase{$1$} 

\hop{$1$ (PSIN)} Transmission over this hop in phase $1$ is accomplished in the same way as in phase $1$ of the scheme that was proposed for the $3$-user $3$-hop X network in \cref{Sec:3user3hop}. The only difference here is the destination scheduling. In particular, here all three destinations are scheduled per time slot as opposed to the single destination scheduling in \cref{Sec:3user3hop}. Eventually, this yields a higher order-$2$ efficiency compared to the single destination scheduling as we will see later (cf. observation \ref{Item:O2efficiency} above). Specifically, $18$ information symbols (six symbols for each destination node) are transmitted over hop $1$ during seven time slots. Then, relay $V_i$, $1\leq i \leq 3$, obtains the following linear combinations after the partial interference nulling (cf. \cref{Eq:PSIN_Output1,Eq:PSIN_Output2,Eq:PSIN_Output3}).
\begin{align}
\frac{y^{(1)}_i(7)}{h^{(1)}_{i1}(7)}{-}\sum_{t=1}^6 \frac{y^{(1)}_i(t)}{h^{(1)}_{i1}(t)}&{=}\sum_{t=1}^6\left(\frac{h^{(1)}_{i2}(7)}{h^{(1)}_{i1}(7)}{-}\frac{h^{(1)}_{i2}(t)}{h^{(1)}_{i1}(t)}\right)u^{[2]}_t{+}\sum_{t=1}^6\left(\frac{h^{(1)}_{i3}(7)}{h^{(1)}_{i1}(7)}{-}\frac{h^{(1)}_{i3}(t)}{h^{(1)}_{i1}(t)}\right)u^{[3]}_t=L_{i \backslash 1}(\xu^{[2]})+L_{i \backslash 1}(\xu^{[3]}), \nonumber \\
\frac{y^{(1)}_i(7)}{h^{(1)}_{i2}(7)}{-}\sum_{t=1}^6 \frac{y^{(1)}_i(t)}{h^{(1)}_{i2}(t)}&{=}\sum_{t=1}^6\left(\frac{h^{(1)}_{i3}(7)}{h^{(1)}_{i2}(7)}{-}\frac{h^{(1)}_{i3}(t)}{h^{(1)}_{i2}(t)}\right)u^{[3]}_t{+}\sum_{t=1}^6\left(\frac{h^{(1)}_{i1}(7)}{h^{(1)}_{i2}(7)}{-}\frac{h^{(1)}_{i1}(t)}{h^{(1)}_{i2}(t)}\right)u^{[1]}_t=L_{i \backslash 2}(\xu^{[3]})+L_{i \backslash 2}(\xu^{[1]}),\nonumber \\
\frac{y^{(1)}_i(7)}{h^{(1)}_{i3}(7)}{-}\sum_{t=1}^6 \frac{y^{(1)}_i(t)}{h^{(1)}_{i3}(t)}&{=}\sum_{t=1}^6\left(\frac{h^{(1)}_{i1}(7)}{h^{(1)}_{i3}(7)}{-}\frac{h^{(1)}_{i1}(t)}{h^{(1)}_{i3}(t)}\right)u^{[1]}_t{+}\sum_{t=1}^6\left(\frac{h^{(1)}_{i2}(7)}{h^{(1)}_{i3}(7)}{-}\frac{h^{(1)}_{i2}(t)}{h^{(1)}_{i3}(t)}\right)u^{[2]}_t=L_{i \backslash 3}(\xu^{[1]})+L_{i \backslash 3}(\xu^{[2]}), \nonumber
\end{align}
where we denoted the vector of information symbol of $S_k$ by $\xu^{[k]}\Def[u^{[k]}_1,u^{[k]}_2,\cdots,u^{[k]}_6]^T$ and its contribution in the received signal of $V_i$ after nulling the effect of $S_{j}$ by $L_{i \backslash j}(\xu^{[k]})$. More specifically,
\begin{align}
L_{i \backslash j}(\xu^{[k]}) \Def \sum_{t=1}^6\left(\frac{h^{(1)}_{ik}(7)}{h^{(1)}_{ij}(7)}{-}\frac{h^{(1)}_{ik}(t)}{h^{(1)}_{ij}(t)}\right)u^{[k]}_t, \hspace{1cm} 1\leq i,j,k \leq 3,\hspace{5mm} j\neq k.
\end{align}

Similar to phase $1$ of the scheme presented in \cref{Sec:3user3hop}, it can be verified that, for any $1\leq k \leq 3$, the six partial linear combinations $\{L_{i \backslash j}(\xu^{[k]})\}_{i=1}^3$, $j\in \{1,2,3\} \backslash \{k\}$, are linearly independent almost surely. Thus, if these six linear combinations are delivered to $D_k$, it can solve them for $\xu^{[k]}$. This will be done during the rest of transmission scheme. In fact, for an arbitrary $N_1$, $N_1$ symbols are transmitted as above in $T^{(1)}_1=\frac{7N_1}{18}$ time slots and $\frac{9N_1}{18}=\frac{N_1}{2}$ linear combinations are generated at $\{V_i\}_{i=1}^3$.

\hop{$2$ (Generation of Order-$2$ Symbols)} During each time slot of this hop, a specific pair of relays and a specific pair of destination nodes are scheduled. The scheduled relays transmit a pair out of the $N_1/2$ linear combinations generated by the end of hop $1$ that include information symbols of the scheduled pair of destinations. Therefore, this hop takes $T^{(2)}_1=\frac{N_1}{2}\times \frac{1}{2}=\frac{N_1}{4}$ time slots in phase $1$. 

For instance, during the first time slot, $(V_1,V_2)$ and $(D_1,D_2)$ are scheduled, and $V_1$ and $V_2$ transmit $L_{1 \backslash 3}(\xu^{[1]})+L_{1\backslash 3}(\xu^{[2]})$ and $L_{2 \backslash 3}(\xu^{[1]})+L_{2 \backslash 3}(\xu^{[2]})$, respectively, while $V_3$ is silent. By the end of this time slot, $D_1$ and $D_2$ receive
\begin{align}
y^{(2)}_1(1)&=h^{(2)}_{11}(1)\left(L_{1 \backslash 3}(\xu^{[1]})+L_{1\backslash 3}(\xu^{[2]})\right)+h^{(2)}_{12}(1)\left(L_{2 \backslash 3}(\xu^{[1]})+L_{2 \backslash 3}(\xu^{[2]})\right) \nonumber \\
&=L'_{1\backslash 3}(\xu^{[1]})+L'_{1\backslash 3}(\xu^{[2]}),\\
y^{(2)}_2(1)&=h^{(2)}_{21}(1)\left(L_{1 \backslash 3}(\xu^{[1]})+L_{1\backslash 3}(\xu^{[2]})\right)+h^{(2)}_{22}(1)\left(L_{2 \backslash 3}(\xu^{[1]})+L_{2 \backslash 3}(\xu^{[2]})\right) \nonumber \\
&=L'_{2\backslash 3}(\xu^{[1]})+L'_{2\backslash 3}(\xu^{[2]}),
\end{align}
respectively, where
\begin{align}
L'_{j\backslash 3}(\xu^{[k]})\Def\sum_{i=1}^2 h_{ji}^{(2)}(t)L_{i\backslash 3}(\xu^{[k]}), \hspace{1cm} k=1,2. \label{Eq:3user3hop-AF-phase1_2sss}
\end{align}
If we deliver $L'_{1\backslash 3}(\xu^{[2]})$ to $D_1$, it can cancel it to obtain $L'_{1\backslash 3}(\xu^{[1]})$, which is a linear combination desired by $D_1$. On the other hand, $L'_{1\backslash 3}(\xu^{[2]})$ itself is desired by $D_2$. Also, since it is solely in terms of the information symbols of $S_2$ and the past CSI, it can be reconstructed by $S_2$ after this time slot using delayed CSI. Therefore, it can be defined as an order-$2$ symbol $u^{[2|1,2]}\Def L'_{1\backslash 3}(\xu^{[2]})$. The order-$2$ symbol $u^{[1|1,2]}\Def L'_{2\backslash 3}(\xu^{[1]})$ can be defined similarly. We note that if we deliver these two order-$2$ symbols to both $D_1$ and $D_2$, then $D_1$ will be provided with two linearly-independent equations in terms of $L_{1 \backslash 3}(\xu^{[1]})$ and $L_{2 \backslash 3}(\xu^{[1]})$ and, thus, can obtain both of them. Likewise, $D_2$ will be able to obtain $L_{1\backslash 3}(\xu^{[2]})$ and $L_{2 \backslash 3}(\xu^{[2]})$. Since two order-$2$ symbols are generated after each time slot of this hop, a total of $T^{(2)}_1\times2=\frac{N_1}{2}$ order-$2$ symbols are generated using different pairs of relays and destination nodes. 
\begin{remark}
This scheme attains the maximal order-$2$ efficiency of $1$. Indeed, each of the generated order-$2$ symbols provides one desired linear combination for each of its intended destinations.
\end{remark}
\begin{remark}
\label{Remark:O2_3user2hop}
For each pair of destination nodes $(D_i,D_j)$, the scheme generates order-$2$ symbols only of type $u^{[i|i,j]}$ and $u^{[j|i,j]}$. In summary, after this phase, it only remains to deliver the $\frac{N_1}{2}$ generated order-$2$ symbols $\{u_\ell^{[1|1,2]},u_\ell^{[2|1,2]},u_\ell^{[2|2,3]},u_\ell^{[3|2,3]},u_\ell^{[1|3,1]},u_\ell^{[3|3,1]}\}_{\ell=1}^{N_1/12}$ to their respective pairs of destination nodes.
\end{remark}

\phase{$2$}

The goal of this phase is to deliver the order-$2$ symbols generated by the end of phase $1$ to their respective pairs of destination nodes. Similar to phase $2$ of the scheme proposed in \cref{Sec:3user3hop}, hop $1$ offloads the order-$2$ symbols to the relays. Thereafter, the relays are responsible for delivering the order-$2$ symbols to the destination nodes without further involvement of the source nodes.

\hop{$1$ (Symbol Offloading)} The main difference between the symbol offloading in this scheme and that in \cref{Sec:3user3hop} is that here only two source nodes, rather than three, transmit per time slot. The reason is that, as emphasized in \cref{Remark:O2_3user2hop}, for each pair of destinations, only their respective pair of source nodes have corresponding order-$2$ symbols to transmit. In particular, this hop takes $T^{(1)}_2=\frac{N_1}{2}\times \frac{1}{2}=\frac{N_1}{4}$ time slots to offload the $\frac{N_1}{2}$ order-$2$ symbols to the relays. Each time slot of this phase is dedicated to a pair of destination nodes. During the time slot dedicated to ($D_i$, $D_j$), $u_\ell^{[i|i,j]}$ and $u_\ell^{[j|i,j]}$ are transmitted by $S_i$ and $S_j$, respectively, while the third source node is silent. During this time slot, each relay receives a linear combination of the two transmitted order-$2$ symbols. Specifically, we have 
\begin{align}
y^{(1)}_1(t)&=h^{(1)}_{1i}(t)u_\ell^{[i|i,j]}+h^{(1)}_{1j}(t)u_\ell^{[j|i,j]},\\
y^{(1)}_2(t)&=h^{(1)}_{2i}(t)u_\ell^{[i|i,j]}+h^{(1)}_{2j}(t)u_\ell^{[j|i,j]}, \\
y^{(1)}_3(t)&=h^{(1)}_{3i}(t)u_\ell^{[i|i,j]}+h^{(1)}_{3j}(t)u_\ell^{[j|i,j]},
\end{align}
where $t$ is the corresponding time slot. If two of the above linear combinations, say $y^{(1)}_1(t)$ and $y^{(1)}_2(t)$, are delivered to both $D_i$ and $D_j$, then both nodes will be able to decode both $u_\ell^{[i|i,j]}$ and $u_\ell^{[j|i,j]}$. Therefore, $y^{(1)}_1(t)$ and $y^{(1)}_2(t)$ can be considered as two \emph{new} order-$2$ symbols, which are now available at the relay side (not the source side). Therefore, $\frac{N_1}{2}$ new order-$2$ are generated at the relays. We emphasize here that the symbol offloading was accomplished at two order-$2$ symbols per time slot in this hop. If the source nodes wanted to deliver the same order-$2$ symbols, rather than their transformed versions, to the relays, it would have been accomplished at $9/7$ symbols per time slot, which is the best known achievable DoF for the transmission over a $3\times 3$ X channel with delayed CSIT \cite{Abdoli2013IC-X}. 

\hop{$2$ (PSIN, Generation of Order-$3$ Symbols, and Final Delivery)} Transmission of the new order-$2$ symbols over hop $2$ can be considered as transmission of order-$2$ symbols over a $3\times 3$ X channel, since \emph{each} relay has order-$2$ symbols for \emph{each} destination nodes. This problem has been addressed in \cite{Abdoli2013IC-X}, wherein the authors proposed a two-phase scheme that includes partial interference nulling together with generation of order-$3$ symbols and final delivery, and achieves $9/8$ DoF for transmission of order-$2$ symbols. Therefore, the $\frac{N_1}{2}$ order-$2$ symbols can be delivered to their respective pairs of destination nodes over hop $2$ in $T^{(2)}_2=\frac{N_1}{2}\times \frac{8}{9}=\frac{4N_1}{9}$ time slots. 

\begin{figure}
\centering
\includegraphics[width=\textwidth]{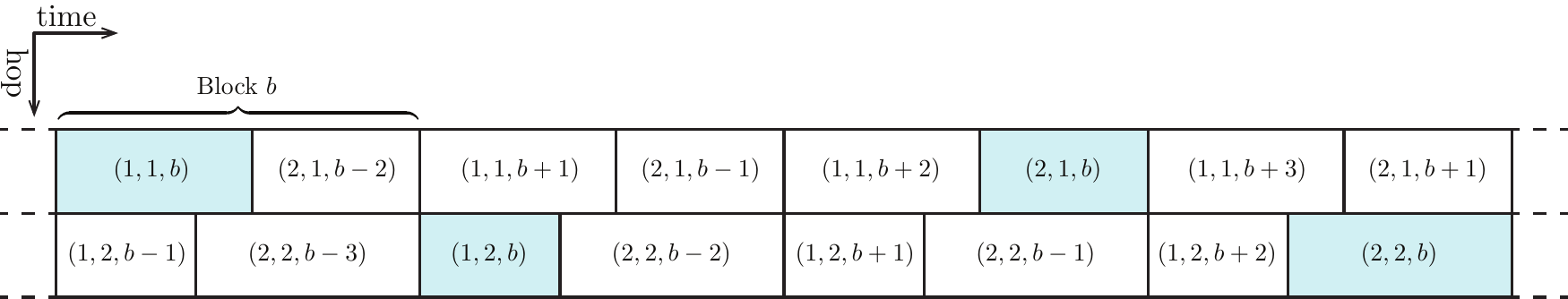}
\caption{Phase-hop interleaver for the $3$-user $2$-hop interference network: The sub-block $(m,k,b)$ denotes transmission in hop $k$ during phase $m$ in round $b$. For any $1\leq b \leq B$, the sub-blocks $(1,1,b)$, $(1,2,b)$, $(2,1,b)$, and $(2,2,b)$ are accomplished in four consecutive blocks.}
\label{Fig:3_User_interleaver}
\end{figure}

Finally, using the phase-hop interleaver of \cref{Fig:3_User_interleaver}, $N_1B$ information symbols are transmitted in $B+3$ blocks. Then, the achieved DoF is equal to 
\begin{align}
\frac{N_1}{\max\{T^{(1)},T^{(2)}\}}=\frac{N_1}{\max\{\frac{7N_1}{18}+\frac{N_1}{4},\frac{N_1}{4}+\frac{4N_1}{9}\}}=\frac{36}{25}.
\end{align}

\subsection{$3$-user $2$-hop Interference Network: Achievability of $16/11$ DoF}
\label{Sec:333_IC_DoF_Improvement}
Following observation \ref{Item:BalancedUtilization} at the beginning of this section, the transmission scheme proposed in \cref{Sec:3_user_2_hop_interference} is not DoF optimal since $T^{(1)}<T^{(2)}$. In this section, we further improve the achievable DoF by modifying phase $2$ of the scheme in order to balance the time slots of hops $1$ and $2$. Specifically, for a fixed $0\leq \beta \leq 1$, for a fraction $\beta$ of the $\frac{N_1}{2}$ order-$2$ symbols, instead of offloading two order-$2$ symbols per time slot, we deliver each order-$2$ symbol to a pair of relay nodes. This is equivalent to the transmission of order-$2$ symbols in the $3\times 3$ X channel with delayed CSIT, which can achieve $9/8$ DoF \cite{Abdoli2013IC-X}. Hence, it takes $\frac{8}{9}\times \frac{\beta N_1}{2}=\frac{4\beta N_1}{9}$ time slots to transmit these order-$2$ symbols over hop $1$.  

Subsequently, the $\frac{\beta N_1}{2}$ order-$2$ symbols will each be available at a pair of relay nodes. As a result, the transmission of these order-$2$ symbols in hop $2$ can be accomplished as in the $3$-user MISO broadcast channel with two antennas at the transmitter and with delayed CSIT. From \cite{maddah2012completely}, we know that this channel has $6/5$ DoF in the transmission of order-$2$ symbols. Therefore, it takes $\frac{5}{6}\times \frac{\beta N_1}{2}= \frac{5\beta N_1}{12}$ time slots in hop $2$ to deliver these order-$2$ symbols to their respective pairs of destination nodes. Now, since the remaining $\frac{(1-\beta)N_1}{2}$ order-$2$ symbols are transmitted over hops $1$ and $2$ as in the original scheme proposed in \cref{Sec:3_user_2_hop_interference}, the total duration of hops $1$ and $2$ is equal to $T^{(1)}=\frac{7N_1}{18}+\frac{4\beta N_1}{9}+\frac{(1-\beta)N_1}{4}$ and $T^{(2)}=\frac{N_1}{4}+\frac{5\beta N_1}{12}+\frac{4(1-\beta)N_1}{9}$, respectively. The optimum value of $\beta$, denoted as $\beta^*$, is obtained by requiring hops $1$ and $2$ to have the same duration or, equivalently, solving the equation
\begin{align}
\frac{7}{18}+\frac{4\beta}{9}+\frac{1-\beta}{4}=\frac{1}{4}+\frac{5\beta}{12}+\frac{4(1-\beta)}{9},
\end{align}
which yields $\beta^*=\frac{1}{4}$. Therefore, we have $T^{(1)}=T^{(2)}=\frac{11}{16}N_1$. The achieved DoF is equal to $\frac{16}{11}$, using the interleaver of \cref{Fig:3_User_interleaver}.

\begin{remark}
The main idea in the improved scheme of \cref{Sec:333_IC_DoF_Improvement} was to spend more time slots in hop $1$ for the transmission of a fraction of the order-$2$ symbols. These extra time slots were utilized towards delivering each of these order-$2$ symbols to a pair of relays rather than a single relay. This, in turn, provided hop $2$ with relay cooperation to deliver the mentioned order-$2$ symbols to their respective pairs of destinations. This relay cooperation yielded a reduction in the time duration of hop $2$. By making a balance between the extra time slots of hop $1$ and the time slot reduction of hop $2$, both hops were forced to have the same duration.
\end{remark}

\section{Concluding Remarks}
\label{Sec:Conclusions}

The impact of multi-hopping on the DoF of interference networks with delayed CSI was investigated in this paper. For the $K$-user $2K$-hop interference network, a multi-phase transmission scheme was proposed that systematically exploited the layered structure of the network and delayed CSI. The achievable DoF of the proposed scheme was shown to scale with $K$. This result provided the first example of a network with distributed transmitters and delayed CSI whose DoF scales with the number of users. By further focusing on the $3$-user case and proposing an improved scheme, it was shown that, in general, our transmission scheme for the $K$-user multi-hop network is not DoF optimal.

This paper assumed delayed global knowledge of network CSI at all nodes. Although this assumption can be justified for networks of moderate sizes, acquiring  global CSI even with finite delay becomes cumbersome as $K$ increases. An interesting future research direction is to investigate interference management in multi-hop interference networks under delayed and local CSI assumption (see \cite{vahid2010capacity,aggarwal2011achieving} for local but instantaneous CSI assumption). Also, since this paper demonstrated the DoF scaling in the $K$-user multi-hop interference network when the number of hops is $2K$ or more (cf. \cref{Remark:Mhop}), an open problem is to find the minimum number of hops and relays per hop that are required to achieve DoF scaling in the layered interference networks with delayed CSI. In this regard, we have the following conjecture.
\begin{conjecture}
\label{Conjecture:MultihopDoFsaturation}
The DoF of the $K$-user $M$-hop interference network with delayed CSI and $M=o(K)$ does not scale with $K$.\footnote{When $f_2(K)$ is nonzero, $f_1(K)=o(f_2(K))$ is equivalent to $\lim_{K\to \infty} \frac{f_1(K)}{f_2(K)}=0$.}
\end{conjecture}
Moreover, it can be shown that the gap between our achievable DoF and the best known upper bound, \ie the $K$-user MISO broadcast channel upper bound \cite{maddah2012completely}, scales with the number of users. Thus, the problem of characterizing the DoF scaling rate of this network with delayed CSI remains open. More generally, it is  curious whether or not having more than $2K$ hops helps to achieve a better scaling rate.

Finding a non-trivial DoF upper bound in networks with distributed transmitters and delayed CSI turns out to be a very challenging problem and is still open even in the single-hop case. The idea behind the only upper bound on the DoF of a $K$-user channel with delayed CSI, \ie the $K$-user MISO broadcast channel upper bound \cite{maddah2012completely}, is to enhance the channel to a multiple-input multiple-output (MIMO) physically degraded broadcast channel, the capacity of which is not increased by feedback \cite{gamal1978feedback}. However, this idea cannot be extended to interference networks, since it does not have any counterpart in these networks.

Recently, the transmission over the $2\times 2$ X channel and $3$-user interference channel with delayed CSI and under linear coding assumption was investigated in \cite{Lashgari2013Xchannel,Lashgari2013XchannelArxiv}, wherein new non-trivial DoF upper bounds, \ie $6/5$ DoF for the X channel (which is tight according to \cite{Ghasemi2011Xchannel}) and $9/7$ DoF for the interference channel were provided. Their key idea is to bound the maximum ratio of the dimensions of the received linear subspaces at a pair of receivers, created by distributed transmitters with delayed CSI. More specifically, they showed that if two distributed transmitters employ linear strategies with delayed CSI, the ratio of the dimensions of the received signals cannot exceed $3/2$. With instantaneous CSI, this ratio can be as large as desired and with no CSI, this ratio is always one. They also conjectured that their upper bounds are also valid without the linear coding restriction. Despite this last progress, no non-trivial upper bound has been reported to date for the $K$-user interference channel with $K>3$, even under the linear coding constraint. 

The situation is even more challenging in the multi-hop networks, for which the upper bounds of \cite{Lashgari2013Xchannel} are not valid anymore. For instance, as shown in \cref{Th:3_user_2_hop_IC_DoF}, the $3$-user $2$-hop interference network can achieve $16/11$ ($>9/7$) DoF with delayed CSI. In fact, since the relays can access mixed linear combinations of the symbols, the destination nodes observe an equivalent channel with \emph{mixed} and distributed transmitters. This interaction between the relays together with parallel time resources of the hops adds more complications to the already open and challenging problem of DoF upper bounding for the interference networks with delayed CSI.

\appendices

\section{Proof of Corollary \ref{Cor:DoFScaling}}
\label{App:ScalingProof}
From \cref{Eq:t_1_Def,Eq:t_2_Def}, one can write
\begin{align}
t_1(q,K)&\leq \frac{\Gamma(q^{-1})(K-1)!}{(q-1)\Gamma(K+q^{-1})}=\frac{\Gamma(1+q^{-1})(K-1)!}{(1-q^{-1})\Gamma(K+q^{-1})}, \label{Eq:t_1_Asymptotic}\\
t_2(q,K)&= \frac{Kq+1}{q(q+1)K} (1 + \epsilon_K)\leq \frac{1}{q}(1+\epsilon_K), \label{Eq:t_2_Asymptotic}
\end{align}
where $\epsilon_K>0$ goes to zero as $K\to \infty$. It is known that for any $c \in \bbR$,
\begin{align}
\lim_{K\to \infty}\frac{(K-1)!K^c}{\Gamma(K+c)}=\lim_{K\to \infty}\frac{\Gamma (K)K^c}{\Gamma(K+c)} =1. \label{Eq:Gamma_c}
\end{align}
Therefore, 
\begin{align}
t_1(q,K) \leq \frac{\Gamma(1+q^{-1})}{(1-q^{-1})K^{q^{-1}}}(1+\epsilon'_K), \label{Eq:t_1_Asymptotic1}
\end{align}
where $\epsilon'_K>0$ goes to zero as $K\to \infty$. Moreover, if $q=q(K)\leq K$ such that $\lim_{K\to \infty}q(K)=+\infty$, we have $\lim_{K\to \infty} \Gamma(1+(q(K))^{-1})=1$ and $\lim_{K\to \infty} 1-(q(K))^{-1}=1$.
Then, in view of \cref{Eq:t_1_Asymptotic1}, we get
\begin{align}
t_1(q(K),K)\leq \frac{1}{K^{(q(K))^{-1}}}(1+\epsilon''_K), \label{Eq:t_1-epsilon}
\end{align}
where $\epsilon''_K>0$ goes to zero as $K\to \infty$. Now, if we choose $q(K)=f^{-1}(K)$ with $f(x)=x^x$, we have $q(K)^{q(K)}=K$ or, equivalently, $K^{(q(K))^{-1}}=q(K)$. Therefore, using \cref{Eq:DoF_lowerbound,Eq:t_1-epsilon,Eq:t_2_Asymptotic}, we can write
\begin{align}
\DoF^\textup{IC}(K,2K) \geq f^{-1}(K)\times \frac{1}{2+ \epsilon_K+\epsilon''_K}.
\end{align}
This last inequality together with $\delta_K \Def \frac{\epsilon_K+\epsilon''_K}{2+\epsilon_K+\epsilon''_K}$ completes the proof.

\section{Proof of Inequality \eqref{Eq:Relaxed_Ineq1}}
\label{App:Inequality}
In this appendix, we prove the following inequality.
\begin{align}
T^{(k)} \leq T^{(1)}+T^{(K)}, \quad 2\leq k \leq K-1.
\end{align}

Using \cref{Eq:N_m+1}, one can write
\begin{align}
N_m=N_1\prod _{j=1}^{m-1}\frac{(K-j)((L-1)(j+1)-1)}{(j+1)((L-1)(K-j)+1)},\quad\quad 1\leq m\leq K. \label{Eq:N_m_expansion}
\end{align}
Let us define $\Lambda_{K,L}(j)$ as
\begin{align}
\Lambda_{K,L}(j) &\Def \frac{(K-j)((L-1)(j+1)-1)}{(j+1)((L-1)(K-j)+1)} \nonumber \\
& =\frac{(L-1)(K-j)(j+1)-(K-j)}{(L-1)(K-j)(j+1)+j+1}, \quad 1 \leq j \leq K-1. \label{Eq:Lambda_Function}
\end{align}
Hence, \cref{Eq:TK,Eq:Tk} can be rewritten as
\begin{align}
T^{(K)}&=N_1\sum_{i=1}^{K}\frac{1}{(L-1)(K-i)+1}\prod _{j=1}^{i-1}\Lambda_{K,L}(j),\label{Eq:TK_Lambda}\\
T^{(k)}&=N_1\left[\frac{1}{K(L-1)}\sum_{i=1}^{k-1}\prod _{j=1}^{i-1}\Lambda_{K,L}(j)+\frac{K(L-1)+1}{KL(L-1)}\prod _{j=1}^{k-1}\Lambda_{K,L}(j)+\frac{1}{K}\prod _{j=1}^{k}\Lambda_{K,L}(j)\right], \label{Eq:Tk_Lambda}
\end{align}
It is easily verified from \cref{Eq:Lambda_Function} that $0 < \Lambda_{K,L}(j) < 1$, for any $1\leq j \leq K-1$. Thus, starting from \cref{Eq:Tk_Lambda}, we have
\begin{align}
T^{(k)} &< N_1\left[\frac{1}{K(L-1)}\sum_{i=1}^{k-1}\prod _{j=1}^{i-1}\Lambda_{K,L}(j)+\frac{K(L-1)+1}{KL(L-1)}+\frac{1}{K}\Lambda_{K,L}(1)\right] \nonumber \\
& \stackrel{\textup{(a)}}{\leq} N_1\left[\sum_{i=1}^{k-1}\frac{1}{(L-1)(K-i)+1}\prod _{j=1}^{i-1}\Lambda_{K,L}(j)+\frac{K(L-1)+1}{KL(L-1)}+\frac{1}{K}\Lambda_{K,L}(1)\right] \label{Eq:T^(k)_ineq2}\nonumber \\
&<N_1\sum_{i=1}^{K}\frac{1}{(L-1)(K-i)+1}\prod _{j=1}^{i-1}\Lambda_{K,L}(j) + N_1\left[\frac{K(L-1)+1}{KL(L-1)}+\frac{1}{K}\Lambda_{K,L}(1)\right] \nonumber \\
&= T^{(K)}+T^{(1)},
\end{align}
where (a) follows from the following inequality, which is valid for any $i \geq 1$ and $L\geq 2$.
\begin{align}
K(L-1) \geq (L-1)(K-i)+1.
\end{align}

\section{Closed Form Expression for $T^{(K)}$}
\label{App:TK_ClosedForm}

In this appendix, we show that the normalized time duration $\bar{T}^{(K)}=\frac{1}{N_1}T^{(K)}$, with $T^{(K)}$ given by \cref{Eq:TK}, is equal to
\begin{align}
\bar{T}^{(K)}=\frac{\alpha}{1-\alpha}\left(\frac{\Gamma(\alpha)(K-1)!}{\Gamma(K+\alpha)}-\frac{1}{K}\right), \label{Eq:App_TK}
\end{align}
where $\Gamma(x)$ is the gamma function, and $\alpha= \frac{1}{L-1}$, $L>2$. By simple manipulations, one can write
\begin{align}
\prod _{j=1}^{i-1}\frac{(K-j)((L-1)(j+1)-1)}{(j+1)((L-1)(K-j)+1)}=
\begin{cases}
\frac{K-i+\alpha}{(1-\alpha)(K-i)}\prod_{j=1}^i\frac{1-\alpha j^{-1}}{1+\alpha (K-j)^{-1}}, &1\leq i \leq K-1, \\
\frac{K-\alpha}{(1-\alpha)K}\prod_{j=1}^{K-1}\frac{1-\alpha j^{-1}}{1+\alpha (K-j)^{-1}}, & i=K,
\end{cases}
\end{align}
Therefore, \cref{Eq:TK} can be rewritten as
\begin{align}
\bar{T}^{(K)}= \frac{1}{1-\alpha}\left[\alpha\left( \Psi(K,\alpha)-\frac{1}{K}\right)+\frac{K-\alpha}{K}\prod_{j=1}^{K-1}\frac{1-\alpha j^{-1}}{1+\alpha (K-j)^{-1}}\right], \label{Eq:TK_replacement}
\end{align}
where $\Psi(K,\alpha)$ is defined as
\begin{align}
\Psi(K,\alpha)\Def \sum_{i=0}^{K-1}\frac{1}{K-i}\prod_{j=1}^i\frac{1-\alpha j^{-1}}{1+\alpha (K-j)^{-1}}.
\end{align}

For $1\leq i \leq K-1$, it can be easily verified that
\begin{align}
\prod_{j=1}^i(1-\alpha j^{-1})&=\frac{-(-\alpha)^{\overline{i+1}}}{i!\alpha}, \label{Eq:prod_nom} \\
\prod_{j=1}^i(1+\alpha (K-j)^{-1})&=\frac{(K-i-1)!\alpha^{\overline{K}}}{(K-1)!\alpha^{\overline{K-i}}}, \label{Eq:prod_denom}
\end{align}
where $x^{\overline{n}}$ is the \emph{rising factorial}, which is defined as
\begin{align}
x^{\overline{n}}\Def x(x+1)\cdots(x+n-1).
\end{align}
Hence, combining \cref{Eq:prod_nom,Eq:prod_denom}, we get
\begin{align}
\prod_{j=1}^i\frac{1-\alpha j^{-1}}{1+\alpha (K-j)^{-1}}&=-\frac{1}{\alpha}\binom{K-1}{i}\frac{\alpha^{\overline{K-i}}(-\alpha)^{\overline{i+1}}}{\alpha^{\overline{K}}} \nonumber \\
&\stackrel{\textup{(a)}}{=}-\frac{1}{\alpha}\binom{K-1}{i}\frac{\Gamma(\alpha+K-i)\Gamma(-\alpha+i+1)}{\Gamma(-\alpha)\Gamma(K+\alpha)} \nonumber  \\
&\stackrel{\textup{(b)}}{=}\binom{K-1}{i}\frac{\Gamma(K+1)B(\alpha+K-i,-\alpha+i+1)}{\Gamma(1-\alpha)\Gamma(K+\alpha)} \nonumber \\
&\stackrel{\textup{(c)}}{=}K!\binom{K-1}{i}\frac{B(\alpha+K-i,-\alpha+i+1)}{\Gamma(1-\alpha)\Gamma(K+\alpha)}, \label{Eq:prod_replacement}
\end{align}
where $B(x,y)$ is the beta function, (a) uses the fact that $x^{\overline{n}}=\frac{\Gamma(x+n)}{\Gamma(x)}$ for $x\neq 0,-1,-2,\cdots$, (b) follows from
\begin{align}
\Gamma(x+1)&=x\Gamma(x), \quad\quad \forall x, \label{Eq:Gamma_Recursive}\\
\Gamma(x)\Gamma(y)&=\Gamma(x+y)B(x,y),\quad\quad  \forall x,y>0, \label{Eq:Gamma_Beta}
\end{align}
and (c) follows from the fact that $\Gamma(K+1)=K!$ for any nonnegative integer $K$.

Plugging \cref{Eq:prod_replacement} into $\Psi(K,\alpha)$, we have
\begin{align}
\Psi(K,\alpha)=\frac{K!}{\Gamma(1-\alpha)\Gamma(K+\alpha)}\sum_{i=0}^{K-1}\binom{K-1}{i}\frac{B(\alpha+K-i,-\alpha+i+1)}{K-i}.
\end{align}
By definition, $B(x,y)=\int_{t=0}^1 t^{x-1}(1-t)^{y-1}dt$. Therefore,
\begin{align}
\Psi(K,\alpha)&=\frac{K!}{\Gamma(1-\alpha)\Gamma(K+\alpha)}\sum_{i=0}^{K-1}\binom{K-1}{i}\frac{1}{K-i}\int_{t=0}^1 t^{\alpha+K-i-1}(1-t)^{-\alpha+i}dt \nonumber \\
&=\frac{K!}{\Gamma(1-\alpha)\Gamma(K+\alpha)}\int_{t=0}^1 t^\alpha (1-t)^{-\alpha} \left(\sum_{i=0}^{K-1}\binom{K-1}{i}\frac{1}{K-i} t^{K-1-i}(1-t)^{i}\right)dt \nonumber \\
&\stackrel{\textup{(a)}}{=}\frac{K!}{\Gamma(1-\alpha)\Gamma(K+\alpha)}\int_{t=0}^1 \frac{t^\alpha (1-t)^{-\alpha}(1-(1-t)^K)}{Kt} dt \nonumber \\
&\stackrel{\textup{(b)}}{=}\frac{(K-1)!}{\Gamma(1-\alpha)\Gamma(K+\alpha)}\left(B(\alpha,1-\alpha)-B(\alpha,K+1-\alpha)\right) \nonumber\\
&\stackrel{\textup{(c)}}{=}\frac{\Gamma(\alpha)(K-1)!}{\Gamma(K+\alpha)}-\frac{\Gamma(\alpha)\Gamma(K+1-\alpha)}{K\Gamma(1-\alpha)\Gamma(K+\alpha)}, \label{Eq:Beta_replacement1}
\end{align}
where (a) results from the following identity
\begin{align}
\sum_{i=0}^{K-1}\binom{K-1}{i}\frac{1}{K-i} x^{K-1-i}y^{i} =\frac{(x+y)^K-y^K}{Kx},\hspace{1cm} \forall x,y\in \bbR, \quad x\neq 0,
\end{align}
(b) follows from the definition of beta function, and (c) uses \cref{Eq:Gamma_Beta}.

On the other hand, using \cref{Eq:prod_replacement}, we get
\begin{align}
\frac{K-\alpha}{K}\prod_{j=1}^{K-1}\frac{1-\alpha j^{-1}}{1+\alpha (K-j)^{-1}}&=(K-1)!\frac{(K-\alpha)B(\alpha+1,K-\alpha)}{\Gamma(1-\alpha)\Gamma(K+\alpha)} \nonumber \\
&\stackrel{\textup{(a)}}{=}\frac{(K-\alpha)\Gamma(\alpha+1)\Gamma(K-\alpha)}{K\Gamma(1-\alpha)\Gamma(K+\alpha)} \nonumber \\
&\stackrel{\textup{(b)}}{=}\frac{\alpha\Gamma(\alpha)\Gamma(K+1-\alpha)}{K\Gamma(1-\alpha)\Gamma(K+\alpha)}, \label{Eq:Beta_replacement3}
\end{align}
where (a) uses \cref{Eq:Gamma_Beta}, and (b) is a result of \cref{Eq:Gamma_Recursive}. Finally, \cref{Eq:App_TK} results from combining \cref{Eq:TK_replacement} with \cref{Eq:Beta_replacement1,Eq:Beta_replacement3}.

\bibliographystyle{IEEEtran}
\bibliography{K_user_K_hop}
\end{document}